\documentclass[prd,amsfonts,showpacs,preprintnumbers,nofootinbib,amsmath,amssymb,superscriptaddress,10pt]{revtex4-2}

\usepackage{bm,graphics,graphicx,epsfig,soul,latexsym,hyperref,float,mathrsfs,multirow,comment}
\usepackage[justification=raggedright,singlelinecheck=true,tableposition=top]{caption}
\usepackage{afterpage}
\usepackage[usenames]{xcolor}
\usepackage{cancel}
\usepackage{dblfloatfix}
\usepackage{enumitem}
\usepackage{xspace}
\usepackage{orcidlink}

\newcommand{\black}{\color{black}}

\graphicspath{{figure/}{./}}
\usepackage{ragged2e}
\usepackage[english]{babel}
\usepackage[normalem]{ulem}
\definecolor{darkgreen}{rgb}{0,0.5,0}

\hypersetup{
    bookmarks=true,         
    unicode=false,          
    pdftoolbar=true,        
    pdfmenubar=true,        
    pdffitwindow=false,     
    pdfstartview={FitH},    
    pdftitle={Aligned_spins_phasing},    
    pdfauthor={Author},     
    pdfsubject={Subject},   
    pdfcreator={Creator},   
    pdfproducer={Producer}, 
    pdfkeywords={keyword1} {key2} {key3}, 
    pdfnewwindow=true,      
    colorlinks=true,       
    linkcolor=red,          
    citecolor=black,        
    filecolor=magenta,      
    urlcolor=darkgreen,           
    linktocpage=true
}

\newcommand{\ord}[1]{\mathcal{O} \left( #1 \right)}

\newcommand{\lal}{\texttt{LALSuite}}

\newcommand{\D}{\mathcal{D}}
\newcommand{\ssout}[1]{}

\newcommand{\pyEFPE}{\texttt{pyEFPE}\xspace}

\newcommand{\Ftwo}{\texttt{TaylorF2}\xspace}

\newcommand{\FtwoEcc}{\texttt{TaylorF2Ecc}\xspace}
\newcommand{\PFtwoEcc}{\texttt{PrecTaylorF2Ecc}\xspace}
\allowdisplaybreaks[4]
\newcommand{\uvec}[1]{\bm{\hat{#1}}}

\newcommand{\IITM}{Department of Physics, Indian Institute of Technology Madras, Chennai 600036, India}
\newcommand{\CSGC}{Centre for Strings, Gravitation and Cosmology, Department of Physics, Indian Institute of Technology Madras, Chennai 600036, India}
\newcommand{\CIERA}{Center for Interdisciplinary Exploration and Research in Astrophysics (CIERA), Northwestern University, 1800 Sherman Ave, Evanston, IL 60201, USA}

\allowdisplaybreaks

\begin{document}

\title{Spin precession effects in the phasing formula of eccentric compact binary inspirals {\black up to} the second post-Newtonian order}

\author{Soham Bhattacharyya\orcidlink{0000-0001-8753-7799}}\email{xeonese@gmail.com}
\affiliation{\IITM}
\affiliation{\CSGC}
\author{Omkar Sridhar\orcidlink{0009-0008-3375-1077
}}\email{omkarnm1401@gmail.com}
\affiliation{\IITM}
\affiliation{\CIERA}



\begin{abstract}
    Compact binary systems that emit gravitational waves (GW) are expected to exhibit orbital eccentricity together with generic spin configurations, leading to the precession of the orbital angular momentum, the individual spins, and the orbital plane around the total angular momentum. Although eccentric binaries with aligned spins have been extensively studied, closed-form post-Newtonian (PN) expressions that simultaneously capture both eccentricity and precessing-spin effects have remained unavailable. The presence of eccentricity significantly complicates orbital evolution, since solving the coupled differential equations generally requires numerical integration, thereby slowing waveform generation. In this work, we exploit the separation of timescales between the orbital motion, spin precession, and radiation reaction, and employ the precession-averaging method introduced by \textit{Morras et al. (2025)} to effectively “average away” the explicit time dependence from the spin–orbit and spin–spin dynamics {\black up to} the second PN order. Building on this framework, we derive analytic phasing formulae from the evolution equations for the orbital frequency and eccentricity, treating eccentricity as a small parameter. Closed-form solutions for the eccentricity evolution and gravitational-wave phase are obtained up to the eighth power of the initial eccentricity $e_0$. The TaylorT2 approximant is generalized to incorporate spin-precession effects, and the orbital phase is calculated in both the time and frequency domains. To further enhance applicability, a resummation is performed on the TaylorT2 phasing formula to improve accuracy for moderate to high initial eccentricities. Taken together, these results provide efficient, closed-form phasing expressions that consistently capture the interplay of eccentricity and precessing spins, thereby enabling accurate and computationally tractable GW modeling for data analysis.
\end{abstract}

\date{\today}
\maketitle
\section{Introduction} \label{sec:intro}

The detection of gravitational waves (GWs) from binary black hole (BBH) mergers~\citep{LIGOScientific:2016aoc}, binary neutron star (BNS) mergers~\citep{LIGOScientific:2017vwq,LIGOScientific:2020aai}, and neutron star–black hole (NSBH) mergers~\citep{LIGOScientific:2021qlt} by the ground-based interferometric network of LIGO~\citep{LIGOScientific:2014pky}, Virgo~\citep{VIRGO:2014yos} and KAGRA ~\citep{Akutsu2021} has inaugurated a transformative era in observational astrophysics, providing an unprecedented window into some of the most energetic phenomena in the Universe. Nearly 300 \cite{LIGOScientific:2025hdt,LIGOScientific:2025slb,LIGOScientific:2025yae,LIGOScientific:2025pvj} compact binary coalescence (CBC) events have been detected by the LIGO-Virgo-KAGRA detector network up until the still ongoing fourth observing run (O4). The massive increase in detection sensitivity is owed to the robust technological upgrades to the detector network \cite{LIGOO4Detector:2023wmz, Jia:2024iqe, LIGO:2024kkz, Capote:2024rmo}.

GW detection is performed using the matched filtering technique \citep{Cutler:1994ys, Poisson:1995ef, Krolak:1995md} in which the observed data is cross-correlated with simulated copies of pre-computed waveforms (also known as templates). The reliability of gravitational-wave analyses is contingent upon the degree of consistency between the modeled templates and the true signal embedded within the detector’s noise. Missing physics in a template bank can lead to biases in the estimation of source parameters, and in some cases, can lead to missed detections altogether \cite{Divyajyoti:2023rht, Nitz:2019spj}.


A key part that comes after parameter estimation is the inference of formation channels of neutron stars and stellar-mass black holes. Eccentricity \cite{Kozai:1962zz,BLSKozai02,VanLandingham:2016ccd,LIDOV1962719,Hoang:2017fvh,Samsing:2013kua,Zevin:2021rtf}, spin-precession \cite{Rodriguez:2016vmx,Gerosa:2018wbw,Banerjee:2023ycw}, and the combination of the two \cite{Shaikh:2025tae}, among other observables, are extremely useful for tracing the formation channels of such compact objects \cite{Singh:2025sww}. It is to be noted that the absence of either eccentricity or spin-precession from template banks can lead to significant biases in parameter estimation \cite{Fumagalli:2023hde,Fumagalli:2024gko,Romero-Shaw:2022fbf,Chatziioannou:2014bma,Pratten:2020igi,Favata:2021vhw,Divyajyoti:2025cwq,Tibrewal:2026jci}. Also, since both introduce amplitude modulations in the GW signal, it becomes important to introduce both effects in waveform models in order to un-entangle the degeneracy caused by the presence of the two. {\black Moreover, incorporating spin-precession effects into waveform models enables the extraction of additional binary parameters that are otherwise difficult to constrain in non-precessing systems. In such systems, these parameters tend to be strongly degenerate with others; precession mitigates this issue by causing the orbital plane to evolve in orientation during the inspiral, effectively providing multiple lines of sight to the same source} \cite{Bambi:2020tsh}. Furthermore, inclusion of all known physical effects in waveform models will help in performing precision tests of General Relativity (GR) \cite{Ma:2019rei, Moore:2020rva,Saini:2022igm,Narayan:2023vhm,Bhat:2022amc,Bhat:2024hyb, Shaikh:2024wyn, Roy:2025xih}.

So far, there has been significant progress in the development of eccentric waveform models. One of the first inspiral-merger-ringdown (IMR) models incorporating eccentricity was ENIGMA \cite{Huerta:2017kez,Chen:2020lzc}, followed recently by its spinning counterpart ESIGMA \cite{Paul:2024ujx}. PN inspired or inspiral only models include \cite{Moore:2016qxz,Huerta:2014eca,Moore:2018kvz,Tanay:2019knc,Tiwari:2020hsu,Maurya:2025shc}, effective-one-body (EOB) models include \cite{Taracchini:2012ig,Liu:2019jpg,Nagar:2024dzj,Gamboa:2024hli, Islam:2025llx}, whereas NR-informed models include \cite{Islam:2021mha, Islam:2024rhm, Islam:2024bza}. Similarly, for precessing spins, PN calculations include \cite{Apostolatos:1994abc,Buonanno:2002fy,Kidder:1995zr,Arun:2008kb,Klein:2013qda}, EOB models include \cite{Pan:2013rra,Ossokine:2020kjp,Pompili:2025cdc}, Phenom models include \cite{Khan:2018fmp,Hamilton:2021pkf}, and NR-informed models include \cite{Varma:2019csw}. Meanwhile, progress is beginning to be made in the combined modeling of precession and eccentricity \cite{Gamba:2024cvy,Liu:2023ldr,Arredondo:2024nsl,Klein:2021jtd,Klein:2018ybm,Klein:2010ti,vandeMeent:2017bcc,Ireland:2019tao,Csizmadia:2012wy,Morras:2025nlp,Phukon:2025yva,Chiaramello:2025bhi}.

During O3, \cite{Gayathri:2020coq} claimed the presence of non-zero eccentricity and precession in the merger event GW190521. Other works such as \cite{Romero-Shaw:2022xko, LIGOScientific:2023lpe, Gupte:2024jfe} claim eccentricity in various other events in GWTC-3. Very recently, there were a series of papers, starting from \cite{Morras:2025xfu}, and others \cite{Planas:2025jny,Romero-Shaw:2025vbc,Kacanja:2025kpr,Jan:2025fps, Tiwari:2025fua, Gamba:2025qfg}, which claimed non-zero eccentricity detection in some GW events in GWTC-4. Keeping all of this in mind, in this paper we develop for the first time, closed-form analytical expressions for the phasing of eccentric and generically spinning orbits, while treating eccentricity as a small parameter. 

This paper is an update to both \cite{Moore:2016qxz} and \cite{Sridhar:2024zms}. Compared to the aligned-spin eccentric phasing of \cite{Sridhar:2024zms}, in this paper we generalize the phasing formula to precessing-spins in elliptical orbits. We give the phasing formula in both time and frequency domain {\black up to} 2 PN, and to an accuracy of the eighth power of the initial eccentricity $e_0$. We use inputs from \cite{Morras:2025nlp} to calculate two Taylor approximants: TaylorT2 and TaylorF2 in the time and frequency domain respectively. We find that the phasing formulae are functions of intrinsic orbital parameters like mass ratios, spins, and quadrupole deformability factors. {\black We also estimate the domain of validity of the phasing formula in both time and frequency domain by performing cycles computation and a mismatch analysis with existing waveform implementations on \lal \cite{lalsuite, swiglal}, an algorithm library for GW analysis.}

This paper is organized as follows: In Sec. \ref{sec:orbevol&phasing}, we first present the inputs, that is, the evolution equations for the eccentricity weighted orbital frequency (or the PN parameter $y$) and the time-eccentricity $e_t$ (to be called as eccentricity henceforth), which we denote as $e$ in this paper. We then turn the evolution equations with respect to time as a single evolution equation for the eccentricity $e$ as a function of the PN parameter $y$ using chain rule. This particular equation can only be solved exactly at the Newtonian order for arbitrary eccentricities. However, for higher PN orders, the evolution equation simply becomes untractable by currently available methods. Nevertheless, as shown by \cite{Moore:2016qxz}, if one treats eccentricity as a small parameter, the differential equation (DE) can be solved exactly, and a closed-form solution can be found {\black up to} some order of the initial eccentricity $e_0$. However, the coefficient of each PN order in the eccentricity expanded DE was constant for the non-spinning and even in the aligned spins case. For generic spins, the coefficients themselves are time dependent, leading to further complications in solving the eccentricity evolution DE. To get around it, Refs. \cite{Klein:2021jtd,Morras:2025nlp} showed that the early inspiral binary problem could be treated as a multiple scale analysis (MSA) problem, which is possible since timescales for the orbit, spin-precession, and radiation-reaction can be clearly segregated. This allows the PN coefficients dependent on the spins to be precession-averaged, effectively taking away the time-dependence from the spin dynamics at the radiation-reaction timescale. Hence, these effective constants can now be put back into the eccentricity evolution DE and then solved as an ordinary DE, which we perform. The eccentricity solution as a function of the eccentricity-weighted orbital frequency $y$ is then utilized to find the time to coalescence $t$ as a function of $y$, and consequently, the time-domain orbital phasing $\phi$, or TaylorT2. We then perform a simple resummation of the TaylorT2 phasing. Finally, we use the Shifted Uniform Asymptotics (SUA) method introduced by \cite{Klein:2013qda} to find the frequency domain phasing $\Psi$.

In Sec. \ref{sec:applications}, we first calculate the number of cycles a real GW signal sweeps across a particular detector frequency range. We compare three different spin scenarios (almost aligned, almost anti-aligned, and perpendicular) for three different detector configurations and mass ranges. This we show for two different initial eccentricities. We then calculate the number of cycles swept by the numerically calculated TaylorT2 phase from the evolution equations and compare it with the analytical small-eccentricity expanded, as well as the resummed TaylorT2 phase number of cycles. Through this comparison, we show the domain of validity of our analytic TaylorT2 phase. We show that the $e_0^8$ expanded TaylorT2 phase incurs less than one cycle, compared to the numerically evaluated TaylorT2 phase, for initial eccentricity $e_0$ as high as 0.7. We also show that the resummed TaylorT2 is valid for initial eccentricities as high as 0.8. {\black We then add our newly computed phasing correction to a frequency domain Taylor waveform and perform mismatch studies to ascertain the domain of validity of our \PFtwoEcc with respect to the eccentric circular aligned-spins waveform \FtwoEcc and the numerically computed eccentric spin-precessing waveform \pyEFPE by \cite{Morras:2025nlp}.}

Finally in Sec. \ref{sec:discussions} we draw some conclusions about the key findings and discusses potential directions for future work.

We use the following notations and conventions in this study: We use geometrized units $G = c = 1$ throughout the study unless otherwise specified. We write spin vectors in bold and use boldfaced hat symbol for the angular momentum unit vector. We use dimensionless spin vectors $\bm{\chi}_{A/S}$ for the results which are related to the $\bm{s}_{1/2}$ vectors of \cite{Morras:2025nlp} as follows

\begin{align}
    \bm{s}_{i} & = \frac{\bm{S}_i}{\mu_i} \qquad \text{for} \, i = 1,2
\end{align}
where
$\bm{S}_i = m_i^2 \bm{\chi}_i$, and $\bm{\chi}_{A/S} = \frac{\bm{\chi}_{1} \mp \bm{\chi}_2}{2}$. However, the approximants in \cite{supplemental} have been expressed in  ($\bm{\chi}_1, \bm{\chi}_2, q_1, q_2$) notation for ease of use for future calculations. We also use symmetric/anti-symmetric spin induced quadrupole moment constants as $\kappa_{A/S} = \frac{q_1 \mp q_2}{2}$ in our results, where $q_i$ were defined in \cite{Morras:2025nlp} after Eq. (C6). Furthermore, we replace all the individual masses $m_i$ with $ \nu = \frac{m_1\,m_2}{M^2} $ and $ \delta = \sqrt{1-4\nu} $, where $M$ is the total mass of the binary system. 

\section{Orbital evolution and phasing}\label{sec:orbevol&phasing}

In this section, we first show the various inputs that we use to calculate the eccentricity evolution. Sec. \ref{subsec:ecc-evol} takes the inputs from \cite{Morras:2025nlp} as the time evolution of the modified PN parameter $y$ and the eccentricity $e$. These are represented as a PN series in $y$ with coefficients depending on the dynamic spins starting from the 1.5 PN spin-orbit (SO) interactions and the 2 PN spin-spin (SS) interactions. In sec. \ref{sec:ecc-evol}, we utilize the precession averaging procedure and the chain rule to obtain the evolution equation of eccentricity as a function of the PN parameter $y$, treating eccentricity as a small parameter. In the same subsection we solve the DE order by order with respect to eccentricity following the procedure of \cite{Sridhar:2024zms}, and provide the result in terms of an initial eccentricity $e_0$ and the initial frequency $y_0$. Then in sec. \ref{sec:TaylorT2} we first calculate the time to coalescence as a function of the PN frequency $y$, followed by the TaylorT2 phase. In sec. \ref{subsec:resum}, we perform a simple resummation of the TaylorT2 phase to increase the domain of validity of the time-domain phase. Finally, in sec. \ref{sec:spa} we calculate the frequency domain phase utilizing the SUA formalism.

\subsection{Eccentricity and PN parameter evolution equations}\label{subsec:ecc-evol}

The evolution equations of $y$ and $e$ are given {\black up to} 2 PN for fully spinning and eccentric orbits by

\begin{subequations}
\label{eq:evol_eqns}
\begin{align}
 \D y &= \nu y^9 \left(a_0 + \sum_{n=2}^{4} a_n  y^n \right), \\
 \D e^2 &= - \nu y^8 \left(b_0 + \sum_{n=2}^{4} b_n  y^n \right),
\end{align}
\end{subequations}

where $\D \equiv \frac{M}{\left(1-e^2\right)^{3/2}} \frac{d}{dt}$, $y = \frac{\left( M \omega \right)^{1/3}}{\sqrt{1 - e^2}}$, and the non-spinning (NS) coefficients are given {\black up to} 2 PN by \cite{Morras:2025nlp}.

\begin{subequations}
    \begin{align}
        a_0 =&\; \frac{32}{5} + \frac{28}{5} e^2 \,, \\
	a_2 =&\; -\frac{1486}{105} - \frac{88}{5} \nu + \left( \frac{12296}{105} - \frac{5258}{45}\nu 
		\right) e^2 + \left( \frac{3007}{84} - \frac{244}{9} \nu \right) e^4 \,, \\
	a_3^\mathrm{NS} =&\; \frac{128 \pi}{5} \phi_y \,, \\
	a_4^\mathrm{NS} =&\; \frac{34103}{2835} + \frac{13661}{315}\nu + \frac{944}{45}\nu^2 + \left(-\frac{489191}{1890} - \frac{209729}{630}\nu + \frac{147443}{270}\nu^2\right) e^2 +  \left(\frac{2098919}{7560} - \frac{2928257}{2520}\nu + \frac{34679}{45}\nu^2\right)e^4 \nonumber\\ 
        & + \left(\frac{53881}{2520}-\frac{7357}{90}\nu+\frac{9392}{135}\nu^2\right)e^6  + \frac{1 - \sqrt{1 
		- e^2}}{\sqrt{1 - e^2}} \bigg[ 16 - \frac{32}{5} \nu + \left(266 - \frac{532}{5}\nu \right) 
		e^2 + \left( -\frac{859}{2} + \frac{859}{5} \nu \right) e^4 \nonumber\\
		& + \left(-65 + 26 \nu \right) e^6 
		\bigg] \,,\\
        b_0 =&\; \frac{608}{15} e^2 + \frac{242}{15} e^4 \,, \\
	b_2 =&\; \left( -\frac{1878}{35} - \frac{8168}{45} \nu \right) e^2 + \left( \frac{59834}{105} - 
		\frac{7753}{15} \nu \right) e^4 + \left( \frac{13929}{140} - \frac{3328}{45} \nu \right) 
		e^6 \,, \\
	b_3^\mathrm{NS} =&\; \frac{788 \pi e^2}{3} \phi_e \,, \\
	b_4^\mathrm{NS} =&\; \left(-\frac{949877}{945} + \frac{18763}{21}\nu + \frac{1504}{5}\nu^2\right) e^2 + \left(-\frac{3082783}{1260} - \frac{988423}{420}\nu + \frac{64433}{20}\nu^2\right) e^4 + \bigg(\frac{23289859}{7560} -\frac{13018711}{2520}\nu \nonumber\\
        & + \frac{127411}{45}\nu^2\bigg) e^6  +  \left(\frac{420727}{1680} - \frac{362071}{1260}\nu + \frac{1642}{9}\nu^2\right) e^8 +
        \sqrt{1 - e^2}\bigg[ \left(\frac{2672}{3} - \frac{5344}{15} \nu \right) e^2 + \left( 2321 - \frac{4642}{5} \nu \right) e^4 \nonumber\\
		&+ \left( \frac{565}{3} - \frac{226}{3} \nu \right) e^6\bigg]
    \end{align}
\end{subequations}

where the tail terms are \cite{Morras:2025nlp}

\begin{subequations}
    \begin{align}
        \phi_y =& \left(1 - e^2 \right)^{7/2} \tilde{\phi} \nonumber \\ 
 =&  1 + \frac{97}{32} e^2 + \frac{49}{128} e^4 - \frac{49}{18432} e^6 - 
		\frac{109}{147456} e^8 - \frac{2567}{58982400} e^{10} + \ord{e^{12}} \,,\\
	\phi_e =& \frac{192 \left(1 - e^2 \right)^{9/2}}{985 e^2} \left( \sqrt{1- e^2} \phi - 
		\tilde{\phi} \right) \nonumber \\ 
 =& 1 + \frac{5969}{3940} e^2 + \frac{24217}{189120} e^4 + \frac{623}{4538880} e^6 - 
		\frac{96811}{363110400} e^8 - \frac{5971}{4357324800} e^{10} + \ord{e^{12}}
    \end{align}
\end{subequations}

The spin-orbit (SO) terms are given {\black up to} 2 PN in terms of $\chi_{eff} = \uvec{L} \cdot \bm{s}$  and $\delta\chi = \uvec{L} \cdot \left( \bm{s}_1 - \bm{s}_2 \right)$ by \cite{Morras:2025nlp}

\begin{subequations}
    \begin{align}
        a_3^\mathrm{SO} =& 
    \left(-\frac{752}{15} -138 e^2 -\frac{611}{30} e^4\right) \chi_\mathrm{eff} + \left(-\frac{152}{15} -\frac{154}{15} e^2 + \frac{17}{30} e^4\right) \delta\mu \delta\chi \, , \\
    b_3^\mathrm{SO} =& 
    e^2 \left(-\frac{3272}{9}-\frac{26263}{45}e^2 - \frac{812}{15}e^4\right) \chi_\mathrm{eff} +e^2 \left(-\frac{3328}{45}-\frac{1993}{45}e^2+\frac{23}{15}e^4\right) \delta \mu  \delta \chi
    \end{align}
\end{subequations}
where
\begin{subequations}
    \begin{align}
        \bm{s}_i = & \frac{\bm{S}_i}{\mu_i} \\
        \mu_i = & \frac{m_i}{M}
    \end{align}
\end{subequations}
for $i \in \left\{ 1,2 \right\}$. $m_i$ are the individual masses of the components of the binary and $M$ is the total mass.

The spin-spin (SS) parts of the evolution equations at 2 PN are given in terms of the following function

\begin{align}
	\sigma(a, b, c, a_1 + a_2 q, b_1 + b_2 q, c_1 + c_2 q) =& a \bm{s}^2 + b \left( \uvec{L} 
		\cdot \bm{s} \right)^2 + c \left| \uvec{L} \times \bm{s} \right|^2 \cos 2 \psi \nonumber\\
		&+ \sum_{i=1}^2 \left[ \left(a_1 + a_2 q_i \right) \bm{s}_i^2 + \left( b_1 + b_2 q_i 
		\right) \left( \uvec{L} \cdot \bm{s}_i \right)^2 + \left(c_1 + c_2 q_i \right) \left| 
		\uvec{L} \times \bm{s}_i \right|^2 \cos 2 \psi_i \right] \, ,
\end{align}

where $q_i$ is the quadrupole parameter, defined in such a way that $q_i = 1$ for black holes. The fully-spinning SS coefficients are given by the following \cite{Morras:2025nlp}

\begin{subequations}
    \label{eq:ab4SS_full}
    \begin{align}
a_4^\mathrm{SS} = &
    \sigma \bigg( - \frac{84}{5} - \frac{228}{5} e^2 - \frac{33}{5} e^4, \frac{242}{5} 
	+ \frac{654}{5} e^2 + \frac{381}{20} e^4, - \frac{447}{10} e^2 - \frac{93}{10} e^4, \frac{88}{5} - 16 q + \left(48 - \frac{216}{5} q \right) e^2 + \left( \frac{69}{10} - \frac{63}{10} q \right) e^4, \nonumber\\
	&  - \frac{244}{5} + 48 q + \left(-132 + \frac{648}{5} q \right) 
	e^2 + \left(- \frac{96}{5} + \frac{189}{10} q \right) e^4, \left(1 - q\right) \left( \frac{447}{10} e^2 + \frac{93}{10} e^4 \right) \bigg) \, , \\
b_4^\mathrm{SS} = &
    \sigma \bigg( \frac{2}{3} 
	-\frac{1961}{15} e^2 - \frac{2527}{12}e^4 - \frac{157}{8} e^6, - \frac{2}{3} + 
	\frac{5623}{15} e^2 + \frac{2393}{4} e^4 + \frac{447}{8} e^6, -\frac{5527}{30} e^2 - 
	\frac{10117}{30} e^4 - \frac{5507}{160} e^6, \nonumber\\
	& -\frac{4}{3} + \left( \frac{682}{5} - \frac{1876}{15} q \right) 
	e^2 + \left(\frac{1337}{6} - \frac{595}{3} q \right) e^4 + \left( \frac{83}{4} - 
	\frac{37}{2} q \right) e^6,	\frac{4}{3} + \left( - \frac{5618}{15} + \frac{1876}{5} q 
	\right) e^2 \nonumber\\
	&+ \left(- \frac{1203}{2} + 595 q \right) e^4 + \left( - \frac{225}{4} + \frac{111}{2} q 
	\right) e^6 , \left( \frac{2764}{15} - \frac{921}{5} q \right) e^2 + \left(\frac{1687}{5} - 
	\frac{5056}{15} q \right) e^4 + \left( \frac{551}{16} - \frac{172}{5} q \right) e^6 \bigg)  \, .
    \end{align}
\end{subequations}

\subsection{Eccentricity evolution}\label{sec:ecc-evol}

Eqs. (\ref{eq:evol_eqns}) can be combined to obtain the eccentricity evolution with respect to the PN parameter $y$. The DE cannot be solved for arbitrary eccentricities using current methods and hence needs to be expanded in small eccentricity in order to obtain an analytical solution. The $e^2$ expanded DE has the following form, after precession averaging of the various PN coefficients appearing in the evolution equations (\ref{eq:evol_eqns})

\begin{align}\label{eq:e2yevol}
    \frac{\D e^2}{\D y} &= y^3 c_{\rm SS,circ} + e^2 \left( c_{\rm NS,ecc} + y^2 \, c_{\rm SO,ecc} + y^3 \, c_{\rm SS,ecc} \right) + e^4 \left( \cdots \right) + \cdots + \mathcal{O}\left(e^8\right)\, ,
\end{align}
where
\begin{subequations}
\begin{align}
    c_{\rm SS,circ} &=  \left(
\frac{5 \nu }{12}
-\frac{5}{48}
\right)
\left(\uvec{\bm{L}}\cdot\bm{\chi}_{A,0}\right)^2
-\frac{5\delta}{24}
\left(\uvec{\bm{L}}\cdot\bm{\chi}_{A,0}\right)
\left(\uvec{\bm{L}}\cdot\bm{\chi}_{S,0}\right)
+
\left(
\frac{5}{48}
-\frac{5 \nu }{12}
\right)
\bm{\chi}_{A,0}\cdot\bm{\chi}_{A,0}
\nonumber\\
&+
\frac{5\delta}{24}
\bm{\chi}_{A,0}\cdot\bm{\chi}_{S,0}
-\frac{5}{48}
\left(\uvec{\bm{L}}\cdot\bm{\chi}_{S,0}\right)^2
+
\frac{5}{48}
\bm{\chi}_{S,0}\cdot\bm{\chi}_{S,0}
 \, , \\
   c_{\rm NS,ecc} &= -\frac{19}{3 y} + y \left( \frac{197 \nu }{18}-\frac{2833}{504} \right) - \frac{377 \pi  y^2}{24} + y^3 \left( \frac{833 \nu ^2}{216}-\frac{32537 \nu }{1008}+\frac{26464169}{1524096} \right) \, , \\
   c_{\rm SO,ecc} &= \frac{157\delta}{18}   \uvec{\bm{L}}\cdot\bm{\chi}_{A,0}+\left(\frac{157}{18}-\frac{55 \nu }{9}\right) \uvec{\bm{L}}\cdot \bm{\chi}_{S,0} \, , \\
   c_{\rm SS,ecc} & =  \left(\uvec{\bm{L}}\cdot\bm{\chi}_{A,0}\right)
\left(\uvec{\bm{L}}\cdot\bm{\chi}_{S,0}\right)
\left[
\left(\frac{89 \nu }{2}-\frac{89}{4}\right) \kappa _A
-\frac{137 \delta }{192}
-\frac{89 \delta \kappa _S}{4}
\right]
\nonumber\\
&+\left(\uvec{\bm{L}}\cdot\bm{\chi}_{S,0}\right)^2
\left[
-\frac{89 \delta \kappa _A}{8}
-\frac{21 \nu }{4}
+\left(\frac{89 \nu }{4}-\frac{89}{8}\right) \kappa _S
-\frac{137}{384}
\right]
\nonumber\\
&+ \left(\uvec{\bm{L}}\cdot\bm{\chi}_{A,0}\right)^2
\left[
-\frac{89 \delta \kappa _A}{8}
+\frac{641 \nu }{96}
+\left(\frac{89 \nu }{4}-\frac{89}{8}\right) \kappa _S
-\frac{137}{384}
\right]
\nonumber\\
&+\bm{\chi}_{A,0}\cdot\bm{\chi}_{S,0}
\left[
\left(\frac{89}{12}-\frac{89 \nu }{6}\right) \kappa _A
-\frac{71 \delta }{192}
+\frac{89 \delta \kappa _S}{12}
\right]
\nonumber\\
&+\bm{\chi}_{A,0}\cdot\bm{\chi}_{A,0}
\left[
\frac{89 \delta \kappa _A}{24}
+\frac{261 \nu }{32}
+\left(\frac{89}{24}-\frac{89 \nu }{12}\right) \kappa _S
-\frac{71}{384}
\right]
\nonumber\\
&+\bm{\chi}_{S,0}\cdot\bm{\chi}_{S,0}
\left[
\frac{89 \delta \kappa _A}{24}
-\frac{89 \nu }{12}
+\left(\frac{89}{24}-\frac{89 \nu }{12}\right) \kappa _S
-\frac{71}{384}
\right] \, ,
\end{align}
\end{subequations}

where $\bm{\chi}_{S/A}$ are the symmetric and asymmetric parts of the dimensionless spin vectors $\bm{\chi}_{1/2}$, related by
\begin{align}
   \bm{\chi}_{S/A} & = \frac{\bm{\chi}_1 \pm \bm{\chi}_2}{2}
\end{align}
and the $0$ subscript stands for the value of the vector at an initial time or frequency.

Eq. (\ref{eq:e2yevol}) at order $e^2$ can be solved utilizing the integrating factor method, since it has the following form

\begin{align}\label{eq:IFform}
    \frac{dR(y)}{dy} + P(y) R(y) \,= &\, Q(y)
\end{align}

which has the following solution

\begin{align}\label{eq:gensol}
    R(y) = & e^{ - \int P(y) dy} \left( \int Q(y) e^{\int P(y) dy} dy \right) + C e^{ - \int P(y) dy}
\end{align}

where the integrating factor is $e^{\int P(y) dy}$ and $C$ is a constant of integration that is fixed by enforcing that $R(y_0) = R_0$. In the present case, $R \equiv e^2$. Since we require a solution that is in a PN series, we perform a Taylor series on the PN parameter in Eq. (\ref{eq:gensol}) to obtain the following solution for $e^2(y)$ {\black up to} 2 PN and to $e_0^2$ accuracy

\begin{align} \label{eq:eevol}
    e^2 (y;\,y_0) =  y^4 \, d_{\rm SS,circ} + e_0^2 \left( d_{\rm NS,ecc} + y^3 \, d_{\rm SO,ecc} + y^4 \, d_{\rm SS,ecc} \right) + e_0^4 \left( \cdots \right) + \cdots + \mathcal{O}\left(e_0^8\right)
\end{align}
{\black where $d_{\rm NS,ecc}$ is provided in \cite{Moore_2016} and $d_{\rm SO,ecc}$ is provided in \cite{Sridhar:2024zms}. The precessing coefficients are as follows}-
\begin{subequations}
\begin{align}
d_{\rm SS,circ} &=
\left(
-\frac{5}{496}
+\frac{5 \nu }{124}
\right)
\left(\uvec{\bm{L}}\cdot\bm{\chi}_{A,0}\right)^2
-\frac{5\delta}{248}
\left(\uvec{\bm{L}}\cdot\bm{\chi}_{A,0}\right)
\left(\uvec{\bm{L}}\cdot\bm{\chi}_{S,0}\right)
\nonumber\\
&\quad -\frac{5}{496}
\left(\uvec{\bm{L}}\cdot\bm{\chi}_{S,0}\right)^2
+\left(
\frac{5}{496}
-\frac{5 \nu }{124}
\right)
\bm{\chi}_{A,0}\cdot\bm{\chi}_{A,0}
\nonumber\\
&\quad+\frac{5}{248} \delta\,
\bm{\chi}_{A,0}\cdot\bm{\chi}_{S,0}
+\frac{5}{496}
\bm{\chi}_{S,0}\cdot\bm{\chi}_{S,0}
\nonumber\\
&\quad+\left(\frac{y_0}{y}\right)^{31/3}
\left[
\left(
\frac{5}{496}
-\frac{5 \nu }{124}
\right)
\left(\uvec{\bm{L}}\cdot\bm{\chi}_{A,0}\right)^2
+\frac{5\delta}{248}
\left(\uvec{\bm{L}}\cdot\bm{\chi}_{A,0}\right)
\left(\uvec{\bm{L}}\cdot\bm{\chi}_{S,0}\right)\right.
\nonumber\\
&\quad\left.
+\frac{5}{496}
\left(\uvec{\bm{L}}\cdot\bm{\chi}_{S,0}\right)^2
+\left(
-\frac{5}{496}
+\frac{5 \nu }{124}
\right)
\bm{\chi}_{A,0}\cdot\bm{\chi}_{A,0}
\right.\nonumber\\
&\quad\left.
-\frac{5\delta}{248}
\bm{\chi}_{A,0}\cdot\bm{\chi}_{S,0} -\frac{5}{496}
\bm{\chi}_{S,0}\cdot\bm{\chi}_{S,0}
\right]\,,
\\[1.0ex]
d_{\rm SS,ecc} &=
\left(\frac{y_0}{y}\right)^{50/3}
\biggl[
\left(
-\frac{725}{75392}
+\frac{725\nu}{18848}
\right)
\left(\uvec{\bm{L}}\!\cdot\!\bm{\chi}_{A,0}\right)^2
-\frac{725\delta}{37696}
\left(\uvec{\bm{L}}\!\cdot\!\bm{\chi}_{A,0}\right)
\left(\uvec{\bm{L}}\!\cdot\!\bm{\chi}_{S,0}\right)
\nonumber\\
&\quad -\frac{725}{75392}
\left(\uvec{\bm{L}}\!\cdot\!\bm{\chi}_{S,0}\right)^2
+\left(
\frac{725}{75392}
-\frac{725\nu}{18848}
\right)
\bm{\chi}_{A,0}\!\cdot\!\bm{\chi}_{A,0}
\nonumber\\
&\quad+\frac{725\delta}{37696}
\bm{\chi}_{A,0}\!\cdot\!\bm{\chi}_{S,0}
+\frac{725}{75392}
\bm{\chi}_{S,0}\!\cdot\!\bm{\chi}_{S,0}
\biggr]
\nonumber\\[1.5ex]
&\quad+\left(\frac{y_0}{y}\right)^{19/3}
\biggl[
\left\{
-\frac{1243}{11904}
-\frac{89\delta}{32}\kappa_A
+\frac{5149\nu}{2976}
+\kappa_S\!\left(-\frac{89}{32}+\frac{89\nu}{16}\right)
\right\}
\left(\uvec{\bm{L}}\!\cdot\!\bm{\chi}_{A,0}\right)^2
\nonumber\\
&\quad+\left\{
-\frac{1243\delta}{5952}
-\frac{89\delta}{16}\kappa_S
+\kappa_A\!\left(-\frac{89}{16}+\frac{89\nu}{8}\right)
\right\}
\left(\uvec{\bm{L}}\!\cdot\!\bm{\chi}_{A,0}\right)
\left(\uvec{\bm{L}}\!\cdot\!\bm{\chi}_{S,0}\right)
\nonumber\\
&\quad+\left\{
-\frac{1243}{11904}
-\frac{89\delta}{32}\kappa_A
-\frac{21\nu}{16}
+\kappa_S\!\left(-\frac{89}{32}+\frac{89\nu}{16}\right)
\right\}
\left(\uvec{\bm{L}}\!\cdot\!\bm{\chi}_{S,0}\right)^2
\nonumber\\
&\quad+\left\{
-\frac{123}{3968}
+\frac{89\delta}{96}\kappa_A
+\kappa_S\!\left(\frac{89}{96}-\frac{89\nu}{48}\right)
+\frac{5887\nu}{2976}
\right\}
\bm{\chi}_{A,0}\!\cdot\!\bm{\chi}_{A,0}
\nonumber\\
&\quad+\left\{
-\frac{123\delta}{1984}
+\frac{89\delta}{48}\kappa_S
+\kappa_A\!\left(\frac{89}{48}-\frac{89\nu}{24}\right)
\right\}
\bm{\chi}_{A,0}\!\cdot\!\bm{\chi}_{S,0}
\nonumber\\
&\quad+\left\{
-\frac{123}{3968}
+\frac{89\delta}{96}\kappa_A
+\kappa_S\!\left(\frac{89}{96}-\frac{89\nu}{48}\right)
-\frac{89\nu}{48}
\right\}
\bm{\chi}_{S,0}\!\cdot\!\bm{\chi}_{S,0}
\biggr]
\nonumber\\[1.5ex]
&\quad+\left(\frac{y_0}{y}\right)^{31/3}
\biggl[
\left\{
\frac{13}{114}
+\frac{89\delta}{32}\kappa_A
+\kappa_S\!\left(\frac{89}{32}-\frac{89\nu}{16}\right)
-\frac{1613\nu}{912}
\right\}
\left(\uvec{\bm{L}}\!\cdot\!\bm{\chi}_{A,0}\right)^2
\nonumber\\
&\quad+\left\{
\frac{13\delta}{57}
+\frac{89\delta}{16}\kappa_S
+\kappa_A\!\left(\frac{89}{16}-\frac{89\nu}{8}\right)
\right\}
\left(\uvec{\bm{L}}\!\cdot\!\bm{\chi}_{A,0}\right)
\left(\uvec{\bm{L}}\!\cdot\!\bm{\chi}_{S,0}\right)
\nonumber\\
&\quad+\left\{
\frac{13}{114}
+\frac{89\delta}{32}\kappa_A
+\kappa_S\!\left(\frac{89}{32}-\frac{89\nu}{16}\right)
+\frac{21\nu}{16}
\right\}
\left(\uvec{\bm{L}}\!\cdot\!\bm{\chi}_{S,0}\right)^2
\nonumber\\
&\quad+\left\{
\frac{13}{608}
-\frac{89\delta}{96}\kappa_A
-\frac{1769\nu}{912}
+\kappa_S\!\left(-\frac{89}{96}+\frac{89\nu}{48}\right)
\right\}
\bm{\chi}_{A,0}\!\cdot\!\bm{\chi}_{A,0}
\nonumber\\
&\quad+\left\{
\frac{13\delta}{304}
-\frac{89\delta}{48}\kappa_S
+\kappa_A\!\left(-\frac{89}{48}+\frac{89\nu}{24}\right)
\right\}
\bm{\chi}_{A,0}\!\cdot\!\bm{\chi}_{S,0}
\nonumber\\
&\quad+\left\{
\frac{13}{608}
-\frac{89\delta}{96}\kappa_A
+\frac{89\nu}{48}
+\kappa_S\!\left(-\frac{89}{96}+\frac{89\nu}{48}\right)
\right\}
\bm{\chi}_{S,0}\!\cdot\!\bm{\chi}_{S,0}
\biggr]
\end{align}
\end{subequations}

where $e(y_0) = e_0$ is some initial eccentricity at some reference time or frequency $y_0$. As can be seen from the above solution, and the DE (\ref{eq:e2yevol}), the presence of misaligned spins lead to the presence of some eccentricity even if at some time or frequency $y_0$ the orbit is circular. While radiation reaction sheds most of the eccentricity from the orbit, spin-precession will typically add some eccentricity to the orbit. 

The solution {\black up to} $\mathcal{O}\left(e_0^4\right)$ is then obtained by dividing (\ref{eq:e2yevol}) by $e^2$ and then substituting the $e_0^2$ order solution in the RHS of the DE, taking a Taylor series in the PN parameter {\black up to} the desired PN order, and then finally multiplying the entire RHS by $e^2$ to make a separable DE, which is then solved by separation of variables method. This process is then continued {\black up to} $\mathcal{O}\left(e_0^8\right)$. The solution $e^2(y)$ {\black up to} $\mathcal{O}\left(e_0^8\right)$ is given in \cite{supplemental}.


\subsection{Spin precession effects in TaylorT2}\label{sec:TaylorT2}

Upon inverting and integrating Eq. \ref{eq:evol_eqns}, we find the coalescence time,  to have the following structure-

\begin{align}\label{eq:t-TT2}
    t =  &- \frac{5}{256 \nu y^8} \biggl\{t_{\rm NS,circ} +   y^3 t_{\rm SO,circ} + y^4 t_{\rm SS,circ} + e_0^2  \biggl[ t_{\rm NS,ecc} + y^3 t_{\rm SO,ecc} + y^4 t_{\rm SS,ecc} \biggr] + \mathcal{O}(e_0^{4}) \biggr\}\,,
\end{align}

{\black where $t_{\rm NS,circ}$ and $t_{\rm NS,ecc}$ are provided in \cite{Moore_2016}, $t_{\rm SO,circ}$ in \cite{Blanchet:2013haa}, and $t_{\rm SO,ecc}$ in \cite{Sridhar:2024zms}. Rest of the coefficients are as follows}-

\begin{subequations}\label{eq:t-T2_coeff}
    \begin{align}
        t_{\rm SS,circ} &=  \biggl(\uvec{\bm{L}}\!\cdot\!\bm{\chi}_{A,0}\biggr)^2
\biggl[
-15\delta\kappa_A
+\frac{4489\nu}{496}
+(30\nu-15)\kappa_S
+\frac{223}{1984}
\biggr]
\nonumber\\
&+\biggl(\uvec{\bm{L}}\!\cdot\!\bm{\chi}_{S,0}\biggr)^2
\biggl[
-15\delta\kappa_A
-\frac{19\nu}{2}
+(30\nu-15)\kappa_S
+\frac{223}{1984}
\biggr]
\nonumber\\
&+\biggl(\uvec{\bm{L}}\!\cdot\!\bm{\chi}_{A,0}\biggr)
\biggl(\uvec{\bm{L}}\!\cdot\!\bm{\chi}_{S,0}\biggr)
\biggl[
(60\nu-30)\kappa_A
+\frac{223\delta}{992}
-30\delta\kappa_S
\biggr]
\nonumber\\
&+\bm{\chi}_{A,0}\!\cdot\!\bm{\chi}_{A,0}
\biggl[
5\delta\kappa_A
+\frac{5431\nu}{496}
+(5-10\nu)\kappa_S
-\frac{471}{1984}
\biggr]
\nonumber\\
&+\bm{\chi}_{A,0}\!\cdot\!\bm{\chi}_{S,0}
\biggl[
(10-20\nu)\kappa_A
-\frac{471\delta}{992}
+10\delta\kappa_S
\biggr]
\nonumber\\
&+\bm{\chi}_{S,0}\!\cdot\!\bm{\chi}_{S,0}
\biggl[
5\delta\kappa_A
-10\nu
+(5-10\nu)\kappa_S
-\frac{471}{1984}
\biggr]
\nonumber\\[1.5ex]
&+\biggl(\frac{y_0}{y}\biggr)^{31/3}
\biggl[
\biggl(
\frac{75}{21328}
-\frac{75\nu}{5332}
\biggr)
\biggl(\uvec{\bm{L}}\!\cdot\!\bm{\chi}_{A,0}\biggr)^2 + \frac{75\delta}{10664}
\biggl(\uvec{\bm{L}}\!\cdot\!\bm{\chi}_{A,0}\biggr)
\biggl(\uvec{\bm{L}}\!\cdot\!\bm{\chi}_{S,0}\biggr)
\nonumber\\
&\quad+
\frac{75}{21328}
\biggl(\uvec{\bm{L}}\!\cdot\!\bm{\chi}_{S,0}\biggr)^2
+\biggl(
\frac{75\nu}{5332}
-\frac{75}{21328}
\biggr)
\bm{\chi}_{A,0}\!\cdot\!\bm{\chi}_{A,0}
\nonumber\\
&\quad-\frac{75\delta}{10664}
\bm{\chi}_{A,0}\!\cdot\!\bm{\chi}_{S,0}
-\frac{75}{21328}
\bm{\chi}_{S,0}\!\cdot\!\bm{\chi}_{S,0}
\biggr] \, , \\
        t_{\rm SS,ecc} &= \biggl(\frac{y_0}{y}\biggr)^{31/3}
\biggl\{
\biggl(\uvec{\bm{L}}\cdot\bm{\chi}_{A,0}\biggr)^2
\biggl[
\frac{1335 \delta \kappa _A}{1376}
-\frac{8065 \nu }{13072}
+\biggl(\frac{1335}{1376}-\frac{1335 \nu }{688}\biggr) \kappa _S
+\frac{65}{1634}
\biggr]
\nonumber\\
&+\biggl(\uvec{\bm{L}}\cdot\bm{\chi}_{A,0}\biggr)
\biggl(\uvec{\bm{L}}\cdot\bm{\chi}_{S,0}\biggr)
\biggl[
\biggl(\frac{1335}{688}-\frac{1335 \nu }{344}\biggr)\kappa_A
+\frac{65 \delta }{817}
+\frac{1335 \delta \kappa _S}{688}
\biggr]
\nonumber\\
&+\biggl(\uvec{\bm{L}}\cdot\bm{\chi}_{S,0}\biggr)^2
\biggl[
\frac{1335 \delta \kappa _A}{1376}
+\frac{315 \nu }{688}
+\biggl(\frac{1335}{1376}-\frac{1335 \nu }{688}\biggr)\kappa_S
+\frac{65}{1634}
\biggr]
\nonumber\\
&+\bm{\chi}_{A,0}\cdot\bm{\chi}_{A,0}
\biggl[
-\frac{445 \delta \kappa _A}{1376}
-\frac{8845 \nu }{13072}
+\biggl(\frac{445 \nu }{688}-\frac{445}{1376}\biggr)\kappa_S
+\frac{195}{26144}
\biggr]
\nonumber\\
&+\bm{\chi}_{A,0}\cdot\bm{\chi}_{S,0}
\biggl[
\biggl(\frac{445 \nu }{344}-\frac{445}{688}\biggr)\kappa_A
+\frac{195 \delta }{13072}
-\frac{445 \delta \kappa _S}{688}
\biggr]
\nonumber\\
&+\bm{\chi}_{S,0}\cdot\bm{\chi}_{S,0}
\biggl[
-\frac{445 \delta \kappa _A}{1376}
+\frac{445 \nu }{688}
+\biggl(\frac{445 \nu }{688}-\frac{445}{1376}\biggr)\kappa_S
+\frac{195}{26144}
\biggr]
\biggr\}
\nonumber\\[1.5ex]
&+\biggl(\frac{y_0}{y}\biggr)^{19/3}
\biggl\{
\biggl(\uvec{\bm{L}}\cdot\bm{\chi}_{A,0}\biggr)^2
\biggl[
-\frac{15447 \delta \kappa _A}{992}
+\frac{287207 \nu }{30752}
+\biggl(\frac{15447 \nu }{496}-\frac{15447}{992}\biggr)\kappa_S
+\frac{7603}{123008}
\biggr]
\nonumber\\
&+\biggl(\uvec{\bm{L}}\cdot\bm{\chi}_{A,0}\biggr)
\biggl(\uvec{\bm{L}}\cdot\bm{\chi}_{S,0}\biggr)
\biggl[
\biggl(\frac{15447 \nu }{248}-\frac{15447}{496}\biggr)\kappa_A
+\frac{7603 \delta }{61504}
-\frac{15447 \delta \kappa _S}{496}
\biggr]
\nonumber\\
&+\biggl(\uvec{\bm{L}}\cdot\bm{\chi}_{S,0}\biggr)^2
\biggl[
-\frac{15447 \delta \kappa _A}{992}
-\frac{4755 \nu }{496}
+\biggl(\frac{15447 \nu }{496}-\frac{15447}{992}\biggr)\kappa_S
+\frac{7603}{123008}
\biggr]
\nonumber\\
&+\bm{\chi}_{A,0}\cdot\bm{\chi}_{A,0}
\biggl[
\frac{5149 \delta \kappa _A}{992}
+\frac{351269 \nu }{30752}
+\biggl(\frac{5149}{992}-\frac{5149 \nu }{496}\biggr)\kappa_S
-\frac{32031}{123008}
\biggr]
\nonumber\\
&+\bm{\chi}_{A,0}\cdot\bm{\chi}_{S,0}
\biggl[
\biggl(\frac{5149}{496}-\frac{5149 \nu }{248}\biggr)\kappa_A
-\frac{32031 \delta }{61504}
+\frac{5149 \delta \kappa _S}{496}
\biggr]
\nonumber\\
&+\bm{\chi}_{S,0}\cdot\bm{\chi}_{S,0}
\biggl[
\frac{5149 \delta \kappa _A}{992}
-\frac{5149 \nu }{496}
+\biggl(\frac{5149}{992}-\frac{5149 \nu }{496}\biggr)\kappa_S
-\frac{32031}{123008}
\biggr]
\biggr\}
\nonumber\\[1.5ex]
&+\biggl(\frac{y_0}{y}\biggr)^{50/3}
\biggl\{
\biggl(
\frac{37575}{4674304}
-\frac{37575 \nu }{1168576}
\biggr)
\biggl(\uvec{\bm{L}}\cdot\bm{\chi}_{A,0}\biggr)^2
+\frac{37575 \delta}{2337152}
\biggl(\uvec{\bm{L}}\cdot\bm{\chi}_{A,0}\biggr)
\biggl(\uvec{\bm{L}}\cdot\bm{\chi}_{S,0}\biggr)
\nonumber\\
&+\biggl(
-\frac{37575}{4674304}
+\frac{37575 \nu }{1168576}
\biggr)
\bm{\chi}_{A,0}\cdot\bm{\chi}_{A,0}
-\frac{37575 \delta}{2337152}
\bm{\chi}_{A,0}\cdot\bm{\chi}_{S,0}
\nonumber\\
&+
\frac{37575}{4674304}
\biggl(\uvec{\bm{L}}\cdot\bm{\chi}_{S,0}\biggr)^2
-\frac{37575}{4674304}
\bm{\chi}_{S,0}\cdot\bm{\chi}_{S,0}
\biggr\}.
    \end{align}    
\end{subequations}
Note that these coefficients up to $\mathcal{O}(e_0^8)$ are given in \cite{supplemental}.

The TaylorT2 phasing $\phi$ follows an evolution equation as follows-

\begin{equation} \label{eq: tt2_evol_eqn}
    \frac{d\langle\phi\rangle}{dy} = \frac{d\langle\phi\rangle}{dt}\frac{dt}{dy}=\left(1-e^2\right)^{3/2}{y^{3}\over M}\frac{dt}{dy}\,,
\end{equation}

Eq. \ref{eq: tt2_evol_eqn} can be integrated using $dt/dy$ from the inversion of Eq. \ref{eq:evol_eqns} to obtain the TaylorT2 phasing as a function of $e_0$ and $y$. It takes the following form-

\begin{align}\label{eq:phi-TT2}
    \phi =  &- \frac{1}{32 \nu y^5} \biggl\{\phi_{\rm NS,circ} +   y^3 \phi_{\rm SO,circ} + y^4 \phi_{\rm SS,circ} + e_0^2 \biggl[ \phi_{\rm NS,ecc} + y^3 \phi_{\rm SO,ecc} + y^4 \phi_{\rm SS,ecc} \biggr] + \mathcal{O}(e_0^{4}) \biggr\}\,,
\end{align}

 where $\phi_{\rm NS,circ}$ and $\phi_{\rm NS,ecc}$ are provided in \cite{Moore_2016}, $\phi_{\rm SO,circ}$ in \cite{Blanchet:2013haa}, and $\phi_{\rm SO,ecc}$ in \cite{Sridhar:2024zms}. Rest of the coefficients are as follows-

\begin{subequations}\label{eq:phi-T2_coeff}
    \begin{align}
        \phi_{\rm SS,circ} &=  \biggl(\uvec{\bm{L}}\cdot\bm{\chi}_{A,0}\biggr)^2
\biggl[
-\frac{75 \delta \kappa_A}{2}+
\textcolor{black}{\frac{22145 \nu}{992}}
+\biggl(75\nu-\frac{75}{2}\biggr)\kappa_S
+\frac{1415}{3968}
\biggr]
+\biggl(\uvec{\bm{L}}\cdot\bm{\chi}_{S,0}\biggr)^2
\biggl[
-\frac{75 \delta \kappa_A}{2} \nonumber\\
&\textcolor{black}{-\frac{95\nu}{4}}
+\biggl(75\nu-\frac{75}{2}\biggr)\kappa_S
+\frac{1415}{3968}
\biggr] + \biggl(\uvec{\bm{L}}\cdot\bm{\chi}_{A,0}\biggr)
\biggl(\uvec{\bm{L}}\cdot\bm{\chi}_{S,0}\biggr)
\biggl[
(150\nu-75)\kappa_A
+\frac{1415\delta}{1984}
-75\delta\,\kappa_S
\biggr]
\nonumber\\
&+\bm{\chi}_{A,0}\cdot\bm{\chi}_{A,0}
\biggl[
\frac{25\delta \kappa_A}{2}
\textcolor{black}{\frac{27455 \nu}{992}}
+\biggl(\frac{25}{2}-25\nu\biggr)\kappa_S
-\frac{2655}{3968}
\biggr]
\nonumber\\
&+\bm{\chi}_{A,0}\cdot\bm{\chi}_{S,0}
\biggl[
(25-50\nu)\kappa_A
-\frac{2655\delta}{1984}
+25\delta\,\kappa_S
\biggr]
\nonumber\\
&+\bm{\chi}_{S,0}\cdot\bm{\chi}_{S,0}
\biggl[
\frac{25\delta \kappa_A}{2}
\textcolor{black}{-25\nu}
+\biggl(\frac{25}{2}-25\nu\biggr)\kappa_S
-\frac{2655}{3968}
\biggr] \nonumber \\
&+\biggl(\frac{y_0}{y}\biggr)^{31/3}
\biggl[
\biggl(
\textcolor{black}{\frac{525\nu}{33728}}
-\frac{525}{134912}
\biggr)
\biggl(\uvec{\bm{L}}\cdot\bm{\chi}_{A,0}\biggr)^2 \left.-\frac{525\delta}{67456}
\biggl(\uvec{\bm{L}}\cdot\bm{\chi}_{A,0}\biggr)
\biggl(\uvec{\bm{L}}\cdot\bm{\chi}_{S,0}\biggr)
\right.
\nonumber\\
&\left.
\textcolor{black}{-\frac{525}{134912}}
\biggl(\uvec{\bm{L}}\cdot\bm{\chi}_{S,0}\biggr)^2
\right. \left.
+\biggl(
\frac{525}{134912}-
\textcolor{black}{\frac{525\nu}{33728}}
\biggr)
\bm{\chi}_{A,0}\cdot\bm{\chi}_{A,0}
\right.
\nonumber\\
&\left.
+\frac{525\delta}{67456}
\bm{\chi}_{A,0}\cdot\bm{\chi}_{S,0}+
\right.\left.
\textcolor{black}{\frac{525}{134912}}
\bm{\chi}_{S,0}\cdot\bm{\chi}_{S,0}
\right] \, , \\
        \phi_{\rm SS,ecc} &=  \biggl(\frac{y_0}{y}\biggr)^{31/3}
\biggl\{
\biggl(\uvec{\bm{L}}\!\cdot\!\bm{\chi}_{A,0}\biggr)^2
\biggl[
-\frac{9345 \delta \kappa_A}{8704}
+\textcolor{black}{\frac{56455\nu}{82688}}
+\biggl(\frac{9345\nu}{4352}-\frac{9345}{8704}\biggr)\kappa_S
-\frac{455}{10336}
\biggr]
\nonumber\\
&+\biggl(\uvec{\bm{L}}\!\cdot\!\bm{\chi}_{A,0}\biggr)
\biggl(\uvec{\bm{L}}\!\cdot\!\bm{\chi}_{S,0}\biggr)
\biggl[
\biggl(\frac{9345\nu}{2176}-\frac{9345}{4352}\biggr)\kappa_A
-\frac{455\delta}{5168}
-\frac{9345\delta\kappa_S}{4352}
\biggr]
\nonumber\\
&+\biggl(\uvec{\bm{L}}\!\cdot\!\bm{\chi}_{S,0}\biggr)^2
\biggl[
-\frac{9345 \delta \kappa_A}{8704}
-\textcolor{black}{\frac{2205\nu}{4352}}
+\biggl(\frac{9345\nu}{4352}-\frac{9345}{8704}\biggr)\kappa_S
-\frac{455}{10336}
\biggr]
\nonumber\\
&+\bm{\chi}_{A,0}\!\cdot\!\bm{\chi}_{A,0}
\biggl[
\frac{3115 \delta \kappa_A}{8704}
+\textcolor{black}{\frac{61915\nu}{82688}}
+\biggl(\frac{3115}{8704}-\frac{3115\nu}{4352}\biggr)\kappa_S
-\frac{1365}{165376}
\biggr]
\nonumber\\
&+\bm{\chi}_{A,0}\!\cdot\!\bm{\chi}_{S,0}
\biggl[
\biggl(\frac{3115}{4352}-\frac{3115\nu}{2176}\biggr)\kappa_A
-\frac{1365\delta}{82688}
+\frac{3115\delta\kappa_S}{4352}
\biggr]
\nonumber\\
&+\bm{\chi}_{S,0}\!\cdot\!\bm{\chi}_{S,0}
\biggl[
\frac{3115 \delta \kappa_A}{8704}
-\textcolor{black}{\frac{3115\nu}{4352}}
+\biggl(\frac{3115}{8704}-\frac{3115\nu}{4352}\biggr)\kappa_S
-\frac{1365}{165376}
\biggr]
\biggr\}
\nonumber\\[1.5ex]
&+\biggl(\frac{y_0}{y}\biggr)^{19/3}
\biggl\{
\biggl(\uvec{\bm{L}}\!\cdot\!\bm{\chi}_{A,0}\biggr)^2
\biggl[
-\frac{18015\delta\kappa_A}{5632}
+\textcolor{black}{\frac{28925\nu}{15872}}
+\biggl(\frac{18015\nu}{2816}-\frac{18015}{5632}\biggr)\kappa_S
+\frac{73355}{698368}
\biggr]
\nonumber\\
&+\biggl(\uvec{\bm{L}}\!\cdot\!\bm{\chi}_{A,0}\biggr)
\biggl(\uvec{\bm{L}}\!\cdot\!\bm{\chi}_{S,0}\biggr)
\biggl[
\biggl(\frac{18015\nu}{1408}-\frac{18015}{2816}\biggr)\kappa_A
+\frac{73355\delta}{349184}
-\frac{18015\delta\kappa_S}{2816}
\biggr]
\nonumber\\
&+\biggl(\uvec{\bm{L}}\!\cdot\!\bm{\chi}_{S,0}\biggr)^2
\biggl[
-\frac{18015\delta\kappa_A}{5632}
-\textcolor{black}{\frac{6315\nu}{2816}}
+\biggl(\frac{18015\nu}{2816}-\frac{18015}{5632}\biggr)\kappa_S
+\frac{73355}{698368}
\biggr]
\nonumber\\
&+\bm{\chi}_{A,0}\!\cdot\!\bm{\chi}_{A,0}
\biggl[
\frac{6005\delta\kappa_A}{5632}
+\textcolor{black}{\frac{426445\nu}{174592}}
+\biggl(\frac{6005}{5632}-\frac{6005\nu}{2816}\biggr)\kappa_S
-\frac{54135}{698368}
\biggr]
\nonumber\\
&+\bm{\chi}_{A,0}\!\cdot\!\bm{\chi}_{S,0}
\biggl[
\biggl(\frac{6005}{2816}-\frac{6005\nu}{1408}\biggr)\kappa_A
-\frac{54135\delta}{349184}
+\frac{6005\delta\kappa_S}{2816}
\biggr]
\nonumber\\
&+\bm{\chi}_{S,0}\!\cdot\!\bm{\chi}_{S,0}
\biggl[
\frac{6005\delta\kappa_A}{5632}
-\textcolor{black}{\frac{6005\nu}{2816}}
+\biggl(\frac{6005}{5632}-\frac{6005\nu}{2816}\biggr)\kappa_S
-\frac{54135}{698368}
\biggr]
\biggr\}
\nonumber\\[1.5ex]
&+\biggl(\frac{y_0}{y}\biggr)^{50/3}
\biggl\{
\biggl(
\frac{215775}{31966208}
-\textcolor{black}{\frac{215775\nu}{7991552}}
\biggr)
\biggl(\uvec{\bm{L}}\!\cdot\!\bm{\chi}_{A,0}\biggr)^2
+\frac{215775\delta}{15983104}
\biggl(\uvec{\bm{L}}\!\cdot\!\bm{\chi}_{A,0}\biggr)
\biggl(\uvec{\bm{L}}\!\cdot\!\bm{\chi}_{S,0}\biggr)
\nonumber\\
&+
\frac{215775}{31966208}
\biggl(\uvec{\bm{L}}\!\cdot\!\bm{\chi}_{S,0}\biggr)^2
+\biggl(
-\frac{215775}{31966208}
+\textcolor{black}{\frac{215775\nu}{7991552}}
\biggr)
\bm{\chi}_{A,0}\!\cdot\!\bm{\chi}_{A,0}
\nonumber\\
&-\frac{215775\delta}{15983104}
\bm{\chi}_{A,0}\!\cdot\!\bm{\chi}_{S,0}
-\frac{215775}{31966208}
\bm{\chi}_{S,0}\!\cdot\!\bm{\chi}_{S,0}
\biggr\}.
    \end{align}    
\end{subequations}
Note that these coefficients up to $\mathcal{O}(e_0^8)$ are given in \cite{supplemental}.

\subsection{Resummation of the TaylorT2 phase}\label{subsec:resum}

Since an eccentricity expanded phase will naturally only be valid for a small range of initial eccentricities, we attempt to increase its accuracy by performing a simple resummation, as highlighted in Sec. III D of Ref.~\citep{Henry:2023tka}. To do this, we choose a resummation ansatz of the form-

\begin{equation}\label{eq: resum-ans}
    \phi^{(\rm resum)}_{ans} = y^{-5} (1-e_0^2)^{1/50} \biggl(d + f ~e_0^2 + g ~e_0^4 + h ~e_0^6 + l ~e_0^8\biggr) \,.
\end{equation}
Note that in Eq.~\eqref{eq: resum-ans}, the exponent of the $(1-e_0^2)$ segment was arrived at by experimenting with various powers, and we finally settled with the one-fiftieth power since it proved to accumulate the smallest error in the number of cycles vis-a-vis the numerical solution. {\black There is no particular physical justification for this choice.}

This resummation ansatz is then expanded once again in $e_0$ and compared with the PN version of the phase to determine the ansatz coefficients $d$, $f$, $g$, $h$ and $l$. Fig.~\ref{fig:comparison} includes comparing the resummed analytical expression with numerical phase results. The resummed phase is found to have the following structure

\begin{align}\label{eq:resum-TT2}
    \phi^{\rm (resum)} =  &- \frac{(1-e_0^2)^{1/50}}{32 \nu y^5} \biggl\{\phi^{\rm (resum)}_{\rm NS,circ} +   y^3 \phi^{\rm (resum)}_{\rm SO,circ} + y^4 \phi^{\rm (resum)}_{\rm SS,circ} + e_0^2 \biggl[ \phi^{\rm (resum)}_{\rm NS,ecc} + y^3 \phi^{\rm (resum)}_{\rm SO,ecc} + y^4 \phi^{\rm (resum)}_{\rm SS,ecc} \biggr] + \mathcal{O}(e_0^{4}) \biggr\}\,,
\end{align}

where,

\begin{subequations}\label{eq:resum-T2_coeff}
    \begin{align}
        \phi^{\rm (resum)}_{\rm NS,circ} = & 1+ y^2 \biggl(\frac{55 \nu }{12}+\frac{3715}{1008}\biggr)-10 \pi  y^3 +  y^4 \biggl(\frac{3085 \nu ^2}{144}+\frac{27145 \nu }{1008}+\frac{15293365}{1016064}\biggr)\, , \\
        \phi^{\rm (resum)}_{\rm NS, ecc} = & \frac{1}{50}-\frac{105}{272} \biggl(\frac{y_0}{y}\biggr)^{19/3}+y^2 \biggl[\frac{11 \nu }{120}+\biggl(\frac{y_0}{y}\biggr)^{25/3}\biggl(\frac{6895 \nu }{3264}-\frac{14165}{13056}\biggr)
   +\biggl(\frac{y_0}{y}\biggr)^{19/3} \biggl(\frac{4155 \nu }{896}-\frac{794635}{75264}\biggr)
   \nonumber \\
   &+\frac{743}{10080}\biggr]+y^3 \biggl[-\frac{13195 \pi}{6528}  \biggl(\frac{y_0}{y}\biggr)^{28/3}-\frac{313 \pi}{960}
     \biggl(\frac{y_0}{y}\biggr)^{19/3}-\frac{\pi }{5}\biggr] + y^4 \biggl[\frac{617 \nu ^2}{1440}+\frac{5429 \nu }{10080}\nonumber \\
   &+\biggl(\frac{y_0}{y}\biggr)^{31/3}\biggl(-\frac{635425 \nu ^2}{117504}+\frac{19505 \nu }{6912}+\frac{5966255}{39481344}\biggr)
   +\biggl(\frac{y_0}{y}\biggr)^{25/3} \biggl(-\frac{272845 \nu ^2}{10752}+\frac{23982055 \nu }{338688}\nonumber \\
   &-\frac{2251200955}{75866112}\biggr)
   +\biggl(\frac{y_0}{y}\biggr)^{19/3} \biggl(\frac{3141415 \nu ^2}{76032}-\frac{30486055 \nu }{354816}+\frac{459664885}{178827264}\biggr)
   +\frac{3058673}{10160640}\biggr] \, , \\
        \phi^{\rm (resum)}_{\rm SO,circ} = & \frac{565 \delta}{24}   \,\, \uvec{\bm{L}}\cdot \bm{\chi}_{A,0}+\biggl(\frac{565}{24}-\frac{95 \nu }{6}\biggr) \uvec{\bm{L}}\cdot \bm{\chi}_{S,0} \, , \\
        \phi^{\rm (resum)}_{\rm SO,ecc} = & \frac{113 \delta }{240}  \uvec{\bm{L}}\cdot\bm{\chi}_{A,0}+\biggl(\frac{y_0}{y}\biggr)^{28/3} \biggl[\frac{5495 \delta}{4896}  \uvec{\bm{L}}\cdot\chi
   _{A,0}+\biggl(\frac{5495}{4896}-\frac{1925 \nu }{2448}\biggr) \uvec{\bm{L}}\cdot\bm{\chi}_{S,0}\biggr]\nonumber \\
   &+\biggl(\frac{y_0}{y}\biggr)^{19/3} \biggl[\frac{179 \delta}{72} 
    \uvec{\bm{L}}\cdot\bm{\chi}_{A,0}+\biggl(\frac{1393 \nu }{360}+\frac{179}{72}\biggr) \uvec{\bm{L}}\cdot\bm{\chi}_{S,0}\biggr]+\biggl(\frac{113}{240}-\frac{19 \nu }{60}\biggr) \uvec{\bm{L}}\cdot\chi
   _{S,0} \, , \\
        \phi^{\rm (resum)}_{\rm SS,circ} = & \biggl(\uvec{\bm{L}}\cdot\bm{\chi}_{A,0}\biggr)^2 \biggl[-\frac{75 \delta  \kappa _A}{2}
+{\black \frac{22145 \nu }{992}}
+\biggl(75 \nu -\frac{75}{2}\biggr) \kappa _S+\frac{1415}{3968}\biggr]
+\biggl(\uvec{\bm{L}}\cdot\bm{\chi}_{S,0}\biggr)^2 
\biggl[-\frac{75 \delta  \kappa _A}{2}
-  \nonumber \\
   & {\black \frac{95 \nu }{4}}+\biggl(75 \nu -\frac{75}{2}\biggr)
   \kappa _S+\frac{1415}{3968}\biggr]
+ \biggl( \uvec{\bm{L}}\cdot\bm{\chi}_{A,0} \biggr) 
\biggl( \uvec{\bm{L}}\cdot\bm{\chi}_{S,0} \biggr) 
\biggl[(150 \nu -75) \kappa _A+\frac{1415 \delta }{1984}-75 \delta  \kappa _S\biggr]\nonumber \\
   &+\biggl(\frac{y_0}{y}\biggr)^{31/3} 
\biggl[\biggl(
{\black \frac{525 \nu }{33728}}
-\frac{525}{134912}\biggr) 
\biggl(\uvec{\bm{L}}\cdot\bm{\chi}_{A,0}\biggr)^2
-\frac{525 \delta }{67456} 
\biggl( \uvec{\bm{L}}\cdot\bm{\chi}_{A,0}  \biggr) 
\biggl( \uvec{\bm{L}}\cdot\bm{\chi}_{S,0} \biggr) 
+\biggl(\frac{525}{134912}\nonumber \\
   &-{\black \frac{525 \nu }{33728}}\biggr) 
\bm{\chi}_{A,0}\cdot\bm{\chi}_{A,0}
+\frac{525 \delta}{67456} 
\bm{\chi}_{A,0}\cdot\bm{\chi}_{S,0} 
{\black -\frac{525}{134912}} 
\biggl(\uvec{\bm{L}}\cdot\bm{\chi}_{S,0}\biggr)^2
+\nonumber \\
   &- {\black \frac{525}{134912}}
\bm{\chi}_{S,0}\cdot\bm{\chi}_{S,0}\biggr]
+\bm{\chi}_{A,0}\cdot\bm{\chi}_{A,0} 
\biggl[\frac{25 \delta  \kappa _A}{2}
+{\black \frac{27455 \nu }{992}}
+\biggl(\frac{25}{2}-25 \nu \biggr) \kappa _S-\frac{2655}{3968}\biggr]\nonumber \\
   &+\bm{\chi}_{A,0}\cdot\bm{\chi}_{S,0} 
\biggl[(25-50 \nu ) \kappa _A-\frac{2655 \delta }{1984}+25 \delta  \kappa _S\biggr]
\nonumber \\
   &+\bm{\chi}_{S,0}\cdot\bm{\chi}_{S,0} 
\biggl[\frac{25 \delta  \kappa _A}{2}
+{\black -25 \nu }
+\biggl(\frac{25}{2}-25 \nu \biggr) \kappa _S-\frac{2655}{3968}\biggr]\, , \\           
        \phi^{\rm (resum)}_{\rm SS,ecc} = &  \biggl(\uvec{\bm{L}}\cdot\bm{\chi}_{A,0}\biggr)^2 
\biggl[-\frac{3 \delta  \kappa _A}{4}
+{\black \frac{4429 \nu }{9920}}
+\biggl(\frac{3 \nu }{2}-\frac{3}{4}\biggr) \kappa _S
+\frac{283}{39680}\biggr]
+\biggl(\uvec{\bm{L}}\cdot\bm{\chi}_{S,0}\biggr)^2 
\biggl[-\frac{3 \delta  \kappa _A}{4}
-{\black \frac{19 \nu }{40}}\nonumber \\
   &+\biggl(\frac{3 \nu}{2}-\frac{3}{4}\biggr) \kappa _S
+\frac{283}{39680}\biggr]
+\biggl(\uvec{\bm{L}}\cdot\bm{\chi}_{A,0} \biggr) 
\biggl(\uvec{\bm{L}}\cdot\bm{\chi}_{S,0} \biggr)
\biggl[\biggl(3 \nu -\frac{3}{2}\biggr) \kappa _A
+\frac{283 \delta }{19840}
-\frac{3 \delta  \kappa _S}{2}\biggr]\nonumber \\
   &+\biggl(\frac{y_0}{y}\biggr)^{31/3}
\biggl\{\biggl(\uvec{\bm{L}}\cdot\bm{\chi}_{A,0}\biggr)^2
\biggl[-\frac{9345 \delta \kappa _A}{8704}
+{\black \frac{1750903 \nu }{2563328}}
+\biggl(\frac{9345 \nu }{4352}-\frac{9345}{8704}\biggr) \kappa _S
-\frac{226079}{5126656}\biggr]\nonumber \\
   &+\biggl(\uvec{\bm{L}}\cdot\bm{\chi}_{A,0} \biggr)
\biggl(\uvec{\bm{L}}\cdot\bm{\chi}_{S,0} \biggr)
\biggl[\biggl(\frac{9345 \nu }{2176}-\frac{9345}{4352}\biggr) \kappa _A
-\frac{226079 \delta }{2563328}
-\frac{9345 \delta  \kappa _S}{4352}\biggr] \nonumber \\
   &+\biggl(\uvec{\bm{L}}\cdot\bm{\chi}_{S,0}\biggr)^2
\biggl[-\frac{9345 \delta  \kappa _A}{8704}
-{\black \frac{2205 \nu }{4352}}
+\biggl(\frac{9345 \nu }{4352}-\frac{9345}{8704}\biggr) \kappa _S
-\frac{226079}{5126656}\biggr] \nonumber \\
   &+\bm{\chi}_{A,0}\cdot\bm{\chi}_{A,0}
\biggl[\frac{3115 \delta  \kappa _A}{8704}
+{\black \frac{1918567 \nu }{2563328}}
+\biggl(\frac{3115}{8704}-\frac{3115 \nu }{4352}\biggr) \kappa_S
-\frac{10479}{1281664}\biggr]\nonumber \\
   &+\bm{\chi}_{A,0}\cdot\bm{\chi}_{S,0}
\biggl[\biggl(\frac{3115}{4352}-\frac{3115 \nu }{2176}\biggr) \kappa _A
-\frac{10479 \delta }{640832}
+\frac{3115 \delta  \kappa _S}{4352}\biggr] \nonumber \\
   &+\bm{\chi}_{S,0}\cdot\bm{\chi}_{S,0}
\biggl[\frac{3115 \delta  \kappa _A}{8704}
+{\black -\frac{3115 \nu }{4352}}
+\biggl(\frac{3115}{8704}-\frac{3115 \nu }{4352}\biggr) \kappa _S
-\frac{10479}{1281664}\biggr]
\biggr\} \nonumber \\
&++\biggl(\frac{y_0}{y}\biggr)^{19/3} 
\biggl\{
\biggl(\uvec{\bm{L}}\cdot\bm{\chi}_{A,0}\biggr)^2 
\biggl[-\frac{18015 \delta  \kappa _A}{5632}
+{\black \frac{28925 \nu }{15872}}
+\biggl(\frac{18015 \nu }{2816}-\frac{18015}{5632}\biggr) \kappa _S
+\frac{73355}{698368}\biggr]\nonumber \\
   &+ \biggl( \uvec{\bm{L}}\cdot\bm{\chi}_{A,0} \biggr) 
\biggl( \uvec{\bm{L}}\cdot\bm{\chi}_{S,0} \biggr) 
\biggl[\biggl(\frac{18015 \nu }{1408}-\frac{18015}{2816}\biggr) \kappa _A
+\frac{73355 \delta }{349184}
-\frac{18015 \delta  \kappa _S}{2816}\biggr]\nonumber \\
   &+\biggl(\uvec{\bm{L}}\cdot\bm{\chi}_{S,0}\biggr)^2 
\biggl[-\frac{18015 \delta  \kappa _A}{5632}
-{\black \frac{6315 \nu }{2816}}
+\biggl(\frac{18015 \nu }{2816}-\frac{18015}{5632}\biggr) \kappa _S
+\frac{73355}{698368}\biggr] \nonumber \\
   &+\bm{\chi}_{A,0}\cdot\bm{\chi}_{A,0} 
\biggl[\frac{6005 \delta  \kappa _A}{5632}
+{\black \frac{426445 \nu }{174592}}
+\biggl(\frac{6005}{5632}-\frac{6005 \nu }{2816}\biggr) \kappa _S
-\frac{54135}{698368}\biggr]\nonumber \\
   &+\bm{\chi}_{A,0}\cdot\bm{\chi}_{S,0}
\biggl[\biggl(\frac{6005}{2816}-\frac{6005 \nu }{1408}\biggr) \kappa _A
-\frac{54135 \delta }{349184}
+\frac{6005 \delta  \kappa _S}{2816}\biggr]\nonumber \\
   &+\bm{\chi}_{S,0}\cdot\bm{\chi}_{S,0} 
\biggl[\frac{6005 \delta  \kappa _A}{5632}
+{\black -\frac{6005 \nu }{2816}}
+\biggl(\frac{6005}{5632}-\frac{6005 \nu }{2816}\biggr) \kappa _S
-\frac{54135}{698368}\biggr]
\biggr\} \nonumber \\
   &+\biggl(\frac{y_0}{y}\biggr)^{50/3} 
\biggl[\biggl(\frac{215775}{31966208}
-{\black \frac{215775 \nu }{7991552}}\biggr)
\biggl(\uvec{\bm{L}}\cdot\bm{\chi}_{A,0}\biggr)^2
+\frac{215775 \delta}{15983104}
\biggl(\uvec{\bm{L}}\cdot\bm{\chi}_{A,0} \biggr)
\biggl(\uvec{\bm{L}}\cdot\bm{\chi}_{S,0}\biggr)\nonumber \\
   &+\biggl({\black \frac{215775 \nu }{7991552}}
-\frac{215775}{31966208}\biggr)
\bm{\chi}_{A,0}\cdot\bm{\chi}_{A,0}
-\frac{215775 \delta}{15983104}
\bm{\chi}_{A,0}\cdot\bm{\chi}_{S,0} \nonumber \\
   &+{\black \frac{215775}{31966208}}
\biggl(\uvec{\bm{L}}\cdot\bm{\chi}_{S,0}\biggr)^2
{\black -\frac{215775}{31966208}}
\bm{\chi}_{S,0}\cdot\bm{\chi}_{S,0}
\biggr]\nonumber \\
   &+\bm{\chi}_{A,0}\cdot\bm{\chi}_{A,0}
\biggl[\frac{\delta  \kappa _A}{4}
+{\black \frac{5491 \nu }{9920}}
+\biggl(\frac{1}{4}-\frac{\nu }{2}\biggr) \kappa _S
-\frac{531}{39680}\biggr] \nonumber \\
   &+\bm{\chi}_{A,0}\cdot\bm{\chi}_{S,0}
\biggl[\biggl(\frac{1}{2}-\nu \biggr) \kappa _A
-\frac{531 \delta }{19840}
+\frac{\delta  \kappa _S}{2}\biggr]+\bm{\chi}_{S,0}\cdot\bm{\chi}_{S,0}
\biggl[\frac{\delta  \kappa _A}{4}
+{\black -\frac{\nu}{2}}
+\biggl(\frac{1}{4}-\frac{\nu }{2}\biggr) \kappa _S
-\frac{531}{39680}\biggr]
    \end{align}    
\end{subequations}
Note that these coefficients up to $\mathcal{O}(e_0^8)$ are given in \cite{supplemental}.

\subsection{Spin effects in the Fourier domain phase (SUA)}\label{sec:spa}

From Eq. (108) of \cite{Morras:2025nlp} it is clear that the SUA Fourier domain phasing formula remains unchanged compared to  {\black the stationary phase approximation (SPA)}. Owing to this simplification, {\black as was shown in Sec. VI E of} Ref.~\cite{Moore:2016qxz}, we find the Fourier domain phasing formula to be
\begin{align}
    \Psi & = 2 \pi f t(f) - 2 \phi[t(f)] + 2 \Phi_0 - \frac{\pi}{4}
\end{align}

The SUA phase $\Psi$ was found to have the following structure

\begin{align}
    \Psi & = \frac{3}{128 \nu y^5} \left\{ \Psi_{\rm NS,circ} + y^3 \Psi_{\rm SO,circ} + y^4 \Psi_{\rm SS,circ} +  e_0^2 \left[ \Psi_{\rm NS,ecc} + y^3 \Psi_{\rm SO,ecc} + y^4 \Psi_{\rm SS,ecc} \right] + e_0^4 \left[ \cdots \right] + \cdots + \mathcal{O}\left(e_0^8\right) \right\}
\end{align}
{\black where $\Psi_{\rm NS,circ}$ and $\Psi_{\rm NS,ecc}$ are provided in \cite{Moore_2016}, $\Psi_{\rm SO,circ}$ in \cite{Arun:2008kb}, and $\Psi_{\rm SO,ecc}$ in \cite{Sridhar:2024zms}. Rest of the coefficients are as follows}-
\begin{subequations}
    \begin{align}
        \Psi_{\rm SS,circ} & = \biggl(\uvec{\bm{L}}\cdot\bm{\chi}_{A,0}\biggr)^2 \biggl[-75 \delta  \kappa _A+  {\black \frac{22095 \nu }{496} }+(150 \nu -75) \kappa _S+\frac{1465}{1984}\biggr]+\biggl(\uvec{\bm{L}}\cdot\bm{\chi}
   _{S,0}\biggr)^2 \biggl[-75 \delta  \kappa _A -  {\black \frac{95 \nu }{2} }\nonumber \\
   &+(150 \nu -75) \kappa _S+\frac{1465}{1984}\biggr]+\biggl(\uvec{\bm{L}}\cdot\bm{\chi}_{A,0} \biggr) \biggl(\uvec{\bm{L}}\cdot\bm{\chi}_{S,0} \biggr)
   \biggl[(300 \nu -150) \kappa _A+\frac{1465 \delta }{992}-150 \delta  \kappa _S\biggr]\nonumber \\
   &+\biggl(\frac{y_0}{y}\biggr)^{31/3}
   \biggl[\biggl(\frac{1625}{181288}-  {\black \frac{1625 \nu }{45322}} \biggr) \biggl(\uvec{\bm{L}}\cdot\bm{\chi}_{A,0}\biggr)^2+\frac{1625 \delta}{90644} \biggl(\uvec{\bm{L}}\cdot\bm{\chi}_{A,0} \biggr) \biggl(\uvec{\bm{L}}\cdot\bm{\chi}
   _{S,0}\biggr)+\biggl( {\black \frac{1625 \nu }{45322}}\nonumber \\
   &-\frac{1625}{181288}\biggr) \bm{\chi}_{A,0}\cdot\bm{\chi}_{A,0}-\frac{1625 \delta}{90644}\bm{\chi}_{A,0}\cdot\bm{\chi}
   _{S,0}+  {\black \frac{1625}{181288}} \biggl(\uvec{\bm{L}}\cdot\bm{\chi}_{S,0}\biggr)^2 \nonumber \\
   &- {\black  \frac{1625}{181288}} \bm{\chi}_{S,0}\cdot\bm{\chi}_{S,0}\biggr]+\bm{\chi}_{A,0}\cdot\bm{\chi}_{A,0} \biggl[25 \delta  \kappa _A+  {\black \frac{27505 \nu }{496}} +(25-50 \nu ) \kappa
   _S-\frac{2705}{1984}\biggr]\nonumber \\
   &+\bm{\chi}_{A,0}\cdot\bm{\chi}_{S,0} \biggl[(50-100 \nu ) \kappa _A-\frac{2705 \delta }{992}+50 \delta  \kappa _S\biggr]+\bm{\chi}_{S,0}\cdot\bm{\chi}_{S,0}
   \biggl[25 \delta  \kappa _A  {\black -50 \nu }+(25-50 \nu ) \kappa _S-\frac{2705}{1984}\biggr] \, , \\
   \Psi_{\rm SS,ecc} & = \biggl(\frac{y_0}{y}\biggr)^{31/3} \biggl\{\biggl(\uvec{\bm{L}}\cdot\bm{\chi}_{A,0}\biggr)^2 \biggl[\frac{28925 \delta  \kappa _A}{11696}-  {\black \frac{524225 \nu
   }{333336}}+\biggl(\frac{28925}{11696}-\frac{28925 \nu }{5848}\biggr) \kappa _S+\frac{4225}{41667}\biggr]\nonumber \\
   &+\biggl( \uvec{\bm{L}}\cdot\bm{\chi}_{A,0} \biggr) \biggl(\uvec{\bm{L}}\cdot\bm{\chi}_{S,0} \biggr)
   \biggl[\biggl(\frac{28925}{5848}-\frac{28925 \nu }{2924}\biggr) \kappa _A+\frac{8450 \delta }{41667}+\frac{28925 \delta  \kappa _S}{5848}\biggr]+\biggl(\uvec{\bm{L}}\cdot\bm{\chi}
   _{S,0}\biggr)^2 \biggl[\frac{28925 \delta  \kappa _A}{11696}\nonumber \\
   &+ {\black \frac{6825 \nu }{5848}}+\biggl(\frac{28925}{11696}-\frac{28925 \nu }{5848}\biggr) \kappa
   _S+\frac{4225}{41667}\biggr]+\bm{\chi}_{A,0}\cdot\bm{\chi}_{A,0} \biggl[-\frac{28925 \delta  \kappa _A}{35088}-  {\black \frac{574925 \nu }{333336}}\nonumber \\
   &+\biggl(\frac{28925 \nu
   }{17544}-\frac{28925}{35088}\biggr) \kappa _S+\frac{4225}{222224}\biggr]+\bm{\chi}_{A,0}\cdot\bm{\chi}_{S,0} \biggl[\biggl(\frac{28925 \nu }{8772}-\frac{28925}{17544}\biggr)
   \kappa _A+\frac{4225 \delta }{111112}-\frac{28925 \delta  \kappa _S}{17544}\biggr]\nonumber \\
   &+\bm{\chi}_{S,0}\cdot\bm{\chi}_{S,0} \biggl[-\frac{28925 \delta  \kappa _A}{35088}  {\black +\frac{28925
   \nu }{17544}}+\biggl(\frac{28925 \nu }{17544}-\frac{28925}{35088}\biggr) \kappa _S+\frac{4225}{222224}\biggr]\biggr\}\nonumber \\
   &+\biggl(\frac{y_0}{y}\biggr)^{19/3}
   \biggl\{\biggl(\uvec{\bm{L}}\cdot\bm{\chi}_{A,0}\biggr)^2 \biggl[-\frac{294955 \delta  \kappa _A}{10912}+  {\black \frac{64385155 \nu }{3967008}}+\biggl(\frac{294955 \nu
   }{5456}-\frac{294955}{10912}\biggr) \kappa _S \nonumber \\
   &+\frac{35057425}{174548352}\biggr]+ \biggl(\uvec{\bm{L}}\cdot\bm{\chi}_{A,0} \biggr) \biggl(\uvec{\bm{L}}\cdot\bm{\chi}_{S,0} \biggr) \biggl[\biggl(\frac{294955 \nu
   }{2728}-\frac{294955}{5456}\biggr) \kappa _A+\frac{35057425 \delta }{87274176}-\frac{294955 \delta  \kappa _S}{5456}\biggr]\nonumber \\
   &+\biggl(\uvec{\bm{L}}\cdot\bm{\chi}_{S,0}\biggr)^2
   \biggl[-\frac{294955 \delta  \kappa _A}{10912}- {\black \frac{92935 \nu }{5456}}+\biggl(\frac{294955 \nu }{5456}-\frac{294955}{10912}\biggr) \kappa
   _S+\frac{35057425}{174548352}\biggr]\nonumber \\
   &+\bm{\chi}_{A,0}\cdot\bm{\chi}_{A,0} \biggl[\frac{294955 \delta  \kappa _A}{32736}+ {\black \frac{864463355 \nu
   }{43637088}}+\biggl(\frac{294955}{32736}-\frac{294955 \nu }{16368}\biggr) \kappa _S-\frac{26037775}{58182784}\biggr]\nonumber \\
   &+\bm{\chi}_{A,0}\cdot\bm{\chi}_{S,0}
   \biggl[\biggl(\frac{294955}{16368}-\frac{294955 \nu }{8184}\biggr) \kappa _A-\frac{26037775 \delta }{29091392}+\frac{294955 \delta  \kappa _S}{16368}\biggr]+\bm{\chi}
   _{S,0}\cdot\bm{\chi}_{S,0} \biggl[\frac{294955 \delta  \kappa _A}{32736}  \nonumber \\
   & {\black - \frac{294955 \nu
   }{16368}}+\biggl(\frac{294955}{32736}-\frac{294955 \nu }{16368}\biggr) \kappa
   _S-\frac{26037775}{58182784}\biggr]\biggr\}+\biggl(\frac{y_0}{y}\biggr)^{50/3} \biggl[\biggl(  {\black \frac{77002025 \nu
   }{1331592352} }\nonumber \\
   &-\frac{77002025}{5326369408}\biggr)
   \biggl(\uvec{\bm{L}}\cdot\bm{\chi}_{A,0}\biggr)^2-\frac{77002025 \delta}{2663184704}\biggl(\uvec{\bm{L}}\cdot\bm{\chi}_{A,0} \biggr) \biggl(\uvec{\bm{L}}\cdot\bm{\chi}_{S,0}\biggr)+\biggl(\frac{77002025}{5326369408}- \nonumber \\
   &  {\black \frac{77002025 \nu }{1331592352}}\biggr) \bm{\chi}_{A,0}\cdot\bm{\chi}_{A,0}+\frac{77002025 \delta}{2663184704}\bm{\chi}_{A,0}\cdot\bm{\chi}_{S,0}  \nonumber \\
   &  {\black - \frac{77002025 
   }{5326369408}} \biggl(\uvec{\bm{L}}\cdot\bm{\chi}_{S,0}\biggr)^2+  {\black \frac{77002025
   }{5326369408}}
   \bm{\chi}_{S,0}\cdot\bm{\chi}_{S,0}\biggr]
    \end{align}
\end{subequations}
Note that these coefficients up to $\mathcal{O}(e_0^8)$ are given in \cite{supplemental}.

\section{Comparison with exact eccentricity and implications for GW data analysis}\label{sec:applications}

Having constructed the Taylor approximants and a resummed version of the orbital phase, we begin this section with a discussion on the validity of the eccentricity-expanded and resummed phase over a range of initial eccentricity ($e_0$), and then calculate the contribution of the spinning section of the $\mathcal{O}(e_0^8)$ expanded orbital phase to the number of cycles accumulated by the waveform between an initial and final frequency determined by the frequency bands probed by current and future GW detectors. We highlight these statistics for LIGO, 3G and LISA detection bands, each with their characteristic frequency range $f \in (f_1, f_2)$. Note that the final frequency $f_2$ is calculated as the minimum of frequency of innermost stable circular orbit ($f_{\rm ISCO}$) and the upper frequency limit of the chosen detector, where-

\begin{equation}
    f_{\rm ISCO} = \frac{c^3}{6^{3/2} \pi G M}
 \end{equation}

Since the eccentricity of the binary orbit becomes negligible near the ISCO, we can approximate $y$ as-

\begin{equation}
    y(f_{\rm ISCO}) \approx \sqrt{x_{\rm ISCO}} = \frac{1}{\sqrt{6}}
\end{equation}

The number of cycles accumulated by $\phi(y(f))$ is then given by-

\begin{equation}\label{eq: cycles_eqn}
    N_{\rm cyc} = \frac{1}{\pi}[\phi(y(f_2)) - \phi(y(f_1))]
\end{equation}

\subsection{PN contribution to the number of GW cycles}\label{sec:num_cycles}

Table \ref{tab:table_cycles_freq_bands} describes the number of GW cycles accumulated within various detector bands for two different eccentricities. We consider three spin configurations with respect to the orbital angular momentum unit vector (which we choose to be $\bm{\uvec{L}} = (0.37525, 0, 0.926924)$), such that the spin magnitudes are set as $||\bm{\chi}_1|| = 0.7$ and $||\bm{\chi}_2|| = 0.8$.

\begin{itemize}
    \item {Almost Aligned (AA): $\bm{\chi}_1 = (-0.138268, 0, 0.686208)$, $\bm{\chi}_2 = (0.464856, 0, 0.651083)$}

    \item {Almost Anti-Aligned (AAA): $\bm{\chi}_1 = (-0.0684358, 0, -0.696647)$, $\bm{\chi}_2 = (0.0766773, 0, -0.796317)$}

    \item {Perpendicular: $\bm{\chi}_1 = (0, 0.7, 0)$, $\bm{\chi}_2 = (0, 0.8, 0)$}
\end{itemize}

The $y$ parametrization we utilize here explicitly depends on eccentricity, and therefore it is no longer useful to segregate the accumulated cycles into circular and eccentric bins, the way it was conducted in \cite{Moore:2016qxz} and \cite{Sridhar:2024zms}.

\begin{table}[H]
	\centering
    \resizebox{1 \columnwidth}{!}{
	\begin{tabular}{|c|c|ccc|ccc|ccc|}
		\hline
		\multicolumn{2}{|c|}{Detector} & \multicolumn{3}{c|}{LIGO/Virgo ($1.4 M_\odot + 1.4 M_\odot$)} & \multicolumn{3}{c|}{3G ($50 M_\odot + 50 M_\odot$)} & \multicolumn{3}{c|}{LISA ($10^5 M_\odot + 10^5 M_\odot$)}  \\
		\hline
		\multicolumn{2}{|c|}{Spin setting} & AA & AAA & Perpendicular & AA & AAA & Perpendicular & AA & AAA & Perpendicular  \\
		\hline
		Eccentricity & PN order &\multicolumn{9}{|c|}{cumulative number of cycles} \\
		
		\hline
	\multirow{3}{*}{$0.1$} &	1.5PN & $87.6563$ & $-88.8566$ & $0.0$ & $36.1736$ & $-36.6689$ & $0.0$ & $111.8984$ & $-113.4306$ & $0.0$ \\

	&	2PN & $-4.5384$ & $-4.6776$ & $1.3725$ & $-2.7109$ & $-2.7941$ & $0.8199$ & $-5.4006$ & $-5.5663$ & $1.6332$\\
            \cline{2-11}
        &    Total & $83.1179$ & $-93.5342$ & $1.3725$ & 
$33.4626$ & $-39.4630$ & $0.8199$ & $106.4977$ & $-118.9969$ & $1.6332$ \\
            \hline
		\hline
	\multirow{3}{*}{$0.5$} &	1.5PN & $68.3498$ & $-69.2857$ & $0.0$ & $27.9104$ &  $-28.2926$ & $0.0$ & $87.7370$ & $-88.9384$ & $0.0$ \\
		
	&	2PN & $-3.8980$ & $-4.0176$ & $1.1785$ & 
$-2.2920$ & $-2.3623$ & $0.6929$ & $-4.6843$ & $-4.8280$ & $1.4162$ \\
		\cline{2-11}
        &    Total & $64.4518$ & $-73.3034$ & $1.1785$ & 
$25.6184$ & $-30.6548$ & $0.6929$ & $83.0527$ & $-93.7664$ & $1.4162$ \\
            \hline
        
	\end{tabular} }
    \vspace{0.5em}
	\captionsetup{format=plain}
        \caption{\justifying Contribution of each PN order to the total number of accumulated cycles by the spinning section of the  $e_0^8$ expanded TaylorT2 phase inside the detector's frequency band, for typical eccentric spinning ($||\bm{\chi}_1|| = 0.7; ||\bm{\chi}_2|| = 0.8$) compact binaries observed by current and future detectors, for two different initial eccentricities ($e_0 = 0.1, 0.5$), and for three different directional configurations of the spins with respect to the total angular momentum. We approximate the frequency bands of LIGO/Virgo A+, Einstein Telescope (ET/CE/3G), and LISA with step functions, respectively between $\bigl[10\,\text{Hz},10^3\,\text{Hz}\bigr]$, $\bigl[1\,\text{Hz},10^4\,\text{Hz}\bigr]$ and $\bigl[10^{-4}\,\text{Hz},10^{-1}\,\text{Hz}\bigr]$ (for justification of step function method, see \cite{Blanchet:2013haa}).} \label{tab:table_cycles_freq_bands} 
      
\end{table}


		
		
        
      

We draw the following inferences from Table \ref{tab:table_cycles_freq_bands}

\begin{itemize}
    \item {Almost Aligned (AA): It is evident that the number of cycles accumulated by the spinning part of the phase at 1.5 PN order (while contributing positively overall) decreases as one increases the initial eccentricity. Conversely, the 2 PN section, which negatively contributes to the GW cycles, shows a decrease in the number of cycles taken away from the orbit at higher initial eccentricities. Both of these conclusions match our expectations, given that eccentricity is known to take away from the GW cycles for binaries with positively aligned spins, as has been described in detail in \cite{Sridhar:2024zms}.}

    \item {Almost Anti-Aligned (AAA): The numbers for the negatively aligned configuration can be understood in the context of the corresponding AA cycles calculations. Given that the 1.5 PN spinning segment only contains spin-orbit ($\bm{\uvec{L}} \cdot \bm{\chi}_{1,2} $) contributions, the flipping of spins while keeping the orbital angular momentum vector intact results in a negation in the number of cycles accumulated. However, the 2 PN section, which is dominated by spin-spin ($\bm{\chi}_1 \cdot \bm{\chi}_2$) and spin-orbit-spin-orbit ($ (\bm{\uvec{L}} \cdot \bm{\chi}_{1})(\bm{\uvec{L}} \cdot \bm{\chi}_{2}) $) contributions to the phase, maintains its negative sign and once again shows a decrease in the number of cycles taken away from the orbit at higher initial eccentricities. These conclusions are once more corroborated by \cite{Sridhar:2024zms}.}

    \item {Perpendicular: The spin-orbit-dominated 1.5 PN spinning segment is completely eliminated for the configuration where spins are taken to be exactly perpendicular to $\bm{\uvec{L}}$. To the contrary, the 2 PN section, comprising mainly of quadratic spin terms, continues to contribute to the number of cycles, which predictably decrease at higher eccentricities.}
\end{itemize}

\subsection{Comparison with exact eccentricity TaylorT2}

\begin{figure}[h]
    \centering
    \includegraphics[width=0.8\linewidth]{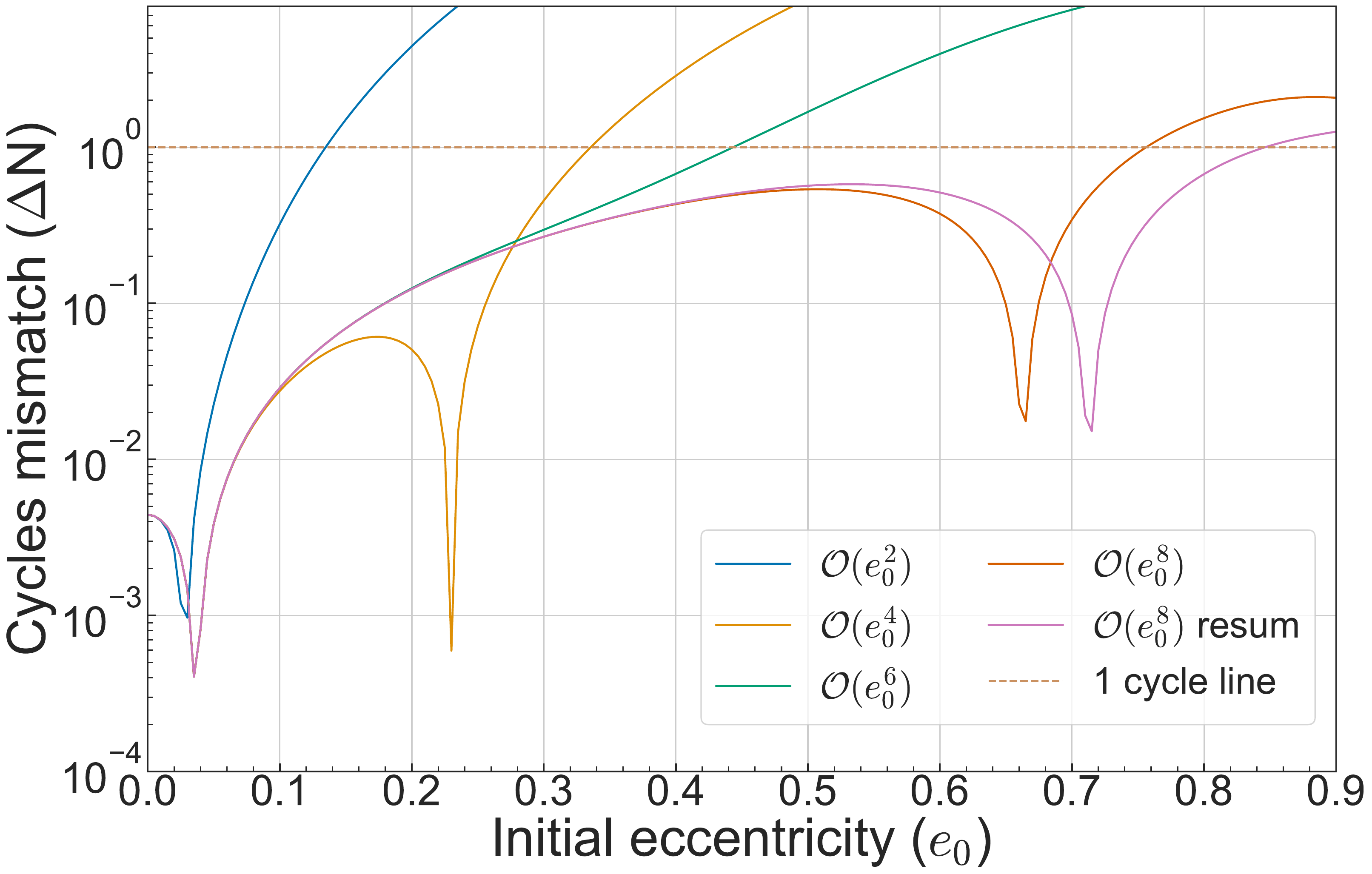}
    \caption{\justifying {\color{black} Absolute value of the} difference in the number of GW cycles between a numerical evolution of the TaylorT2 phase (valid for any initial eccentricity $e_0$) and the analytical TaylorT2 phase in Eqs. (\ref{eq:phi-TT2}) and (\ref{eq:resum-TT2}). The 1-cycle line roughly represents where the phase error between the exact eccentricity solution and the analytical low eccentricity solution becomes significant and our approximation breaks down. The above plot has been made for a system with component masses $m_1 = m_2 = 1.4\,M_\odot$. The magnitude of the individual spins are chosen to be $||\bm{\chi}_1|| = 0.7$, $||\bm{\chi}_2|| = 0.8$, with their orientations being on the x-y plane. Note that the total angular momentum vector is directed along the z-axis.}
    \label{fig:comparison}
\end{figure}

The accuracy of the orbital phase expressions calculated in Sec. \ref{sec:TaylorT2} and Sec. \ref{subsec:resum} can be tested by comparing with an orbital phase computed by numerically integrating the evolution equations derived from \cite{Morras:2025nlp}, for a specific setting of spin ($\bm{\chi}_1$ and $\bm{\chi}_2$) and angular momentum ($\bm{\uvec{L}}$) vectors. In Fig. \ref{fig:comparison}, we plot the difference in the number of cycles between the numerically and analytically calculated orbital phase at various initial eccentricity orders, where, using Eq. (\ref{eq: cycles_eqn})-

\begin{equation}
    \Delta N_{\rm cyc} = N_{\rm cyc, numerical} - N_{\rm cyc, analytical} 
\end{equation}

The gauge we use to quantify the accuracy of our approximants is that of a single GW cycle, beyond which the error becomes significant enough so as to deem the approximant unreliable. For our chosen spin setting, we find for instance that the $\mathcal{O}(e_0^2)$ expression remains reliable only up to an initial eccentricity of 0.14, whereas the $\mathcal{O}(e_0^8)$ expanded orbital phase can demonstrably achieve accuracy all the way to  $e_0 \lesssim 0.75$. Meanwhile, the resummed version extends the validity to $e_0 \lesssim 0.82$. These estimates therefore illustrate the importance of including higher orders of eccentricity while estimating analytical versions of Taylor approximants.

{\black
\subsection{Mismatch computations}
To estimate the effects of the spin-precessing eccentric corrections to the waveform, we first included the newly computed analytic frequency domain phasing to a private branch of \lal. The phasing correction includes the eccentric spin-precessing terms until 2 PN, as well as eccentric aligned-spins terms until 3 PN. We however, include only the Newtonian amplitude to our waveform and restrict the initial eccentricity accuracy {\black up to} $\mathcal{O}(e_0^2)$. Then we used the standard definition of match as noise weighted inner product of two waveforms, one of which contains our newly computed phasing terms. We define the noise weighted inner product as
\begin{equation}\label{eq:inner_product}
    \left(h_1 | h_2\right) \equiv 4 \Re \left[ \int_{f_{start}}^{f_{end}} \frac{\tilde{h}_1(f) \tilde{h}_2^*(f)}{S_n(f)} \right] 
\end{equation}
where $\tilde{h}_1(f)$ and $\tilde{h}_1(f)$ are the Fourier transforms of $h_1$ and $h_2$ respectively, $f_{start}$ and $f_{end}$ are the start and end cutoff frequencies which we take to be $20$ Hz and a fraction of the corresponding Kerr ISCO frequency ($=0.8\times 
f_{Kerr,ISCO}$) respectively, $S_n(f)$ is the power spectral density (PSD), and $*$ denotes complex conjugation. We take $S_n(f)$ to be the advanced LIGO, zero detuned, and high laser power configuration. We define the match as the following \cite{Purrer:2014fza}
\begin{equation}\label{eq:match}
    \mathcal{M}(h_1, h_2) \equiv \, \dfrac{(h_1|h_2)}{\sqrt{(h_1|h_1) (h_2|h_2)}}\,, 
\end{equation}
such that $\mathcal{M}\mathcal{M} = 1-\mathcal{M}$ is the mismatch between the two waveforms $h_1$ and $h_2$. 

We first perform some benchmark studies of our waveform \PFtwoEcc with \FtwoEcc, which contains circular spins and is accurate {\black up to} $\mathcal{O}(e_0^2)$. As a first test, we compute the mismatch of \PFtwoEcc with \FtwoEcc while varying the purely perpendicular initial spin configurations, as is defined in the co-precessing frame with respect to the unit orbital angular momentum vector, and for an equal mass system for two values of eccentricity in Fig. \ref{fig:TF2Ecc_and_PTF2Ecc_mismatch_side_by_side}.

\begin{figure*}[t]
  \centering
  \includegraphics[width=0.49\textwidth]{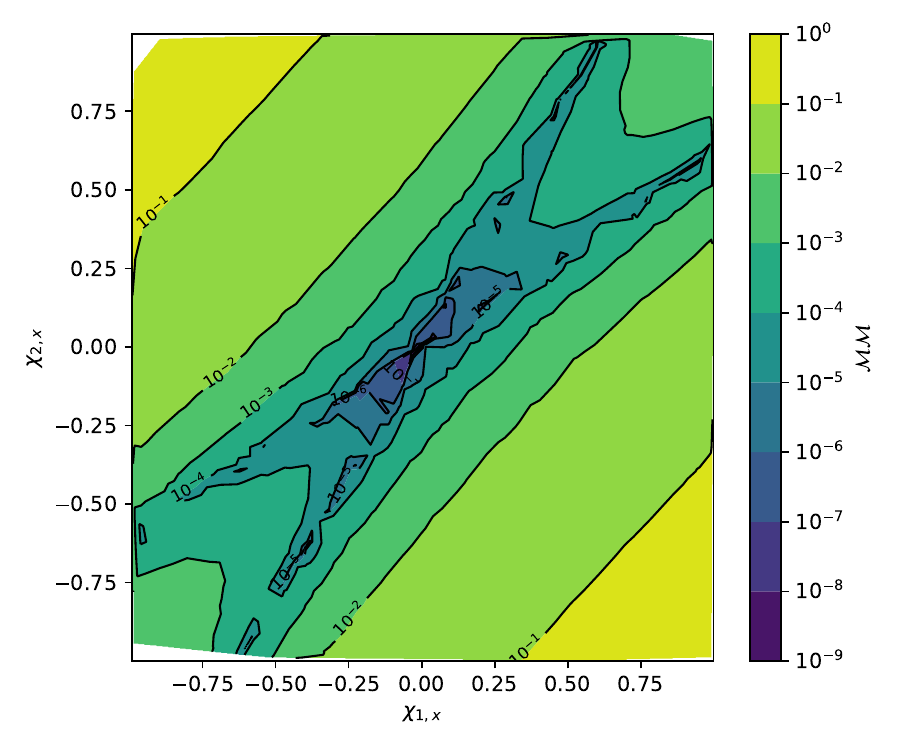}\hfill
  \includegraphics[width=0.49\textwidth]{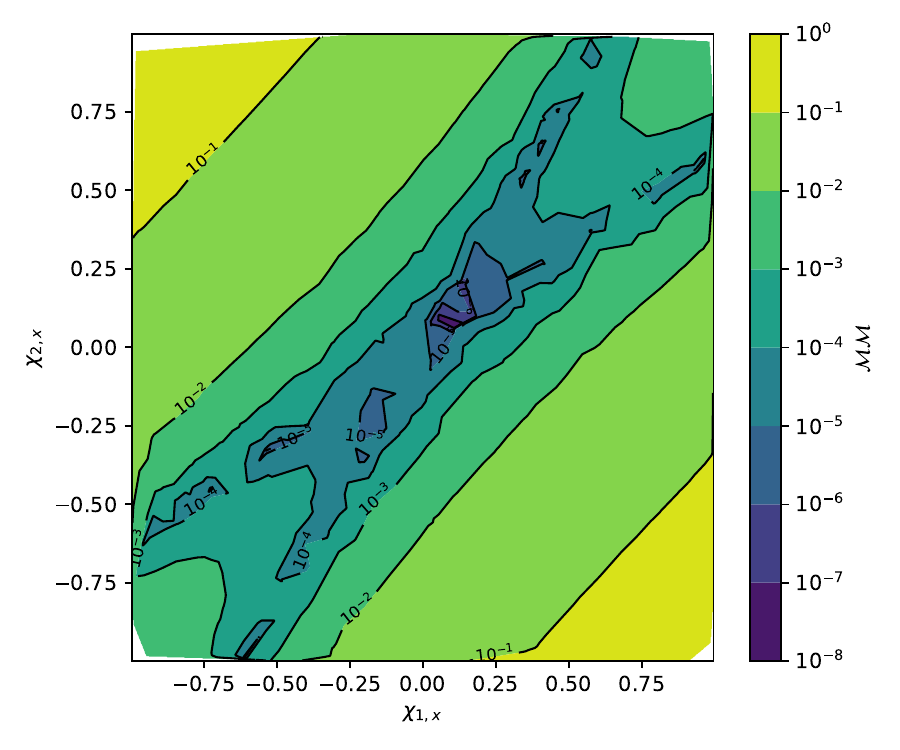}

  \caption{In the above figures, we compute the mismatch between \FtwoEcc and \PFtwoEcc for two different values of initial eccentricity $e_0$. In the left figure we used a contour plot to quantify the mismatch for various initial x component of spins $\chi_{1,x}$ and $\chi_{2,x}$ for zero initial eccentricity. While on the right figure, we plot the same for $e_0 = 0.3$. One notices that while the trend is almost the same for both the figures, however, the center of the figures reveal a slight increase in the mismatch value when the initial eccentricity is increased from zero to 0.3.}
  \label{fig:TF2Ecc_and_PTF2Ecc_mismatch_side_by_side}
\end{figure*}

We then perform a mismatch computation between \PFtwoEcc and \pyEFPE, first by varying the chirp mass ($\mathcal{M}_{chirp}$) and $e_0$ for fixed values of an arbitrary spin configuration in the left plot of Fig. \ref{fig:TF2Ecc_and_pyEFPE_mismatch_side_by_side}, and then by varying the effective spin precession parameter $\chi_p$, as can be found in Ref. \cite{Schmidt:2014iyl}, and $e_0$, for an equal mass and z-component of spins in the right plot of Fig. \ref{fig:TF2Ecc_and_pyEFPE_mismatch_side_by_side}.

\begin{figure*}[t]
  \centering
  \includegraphics[width=0.49\textwidth]{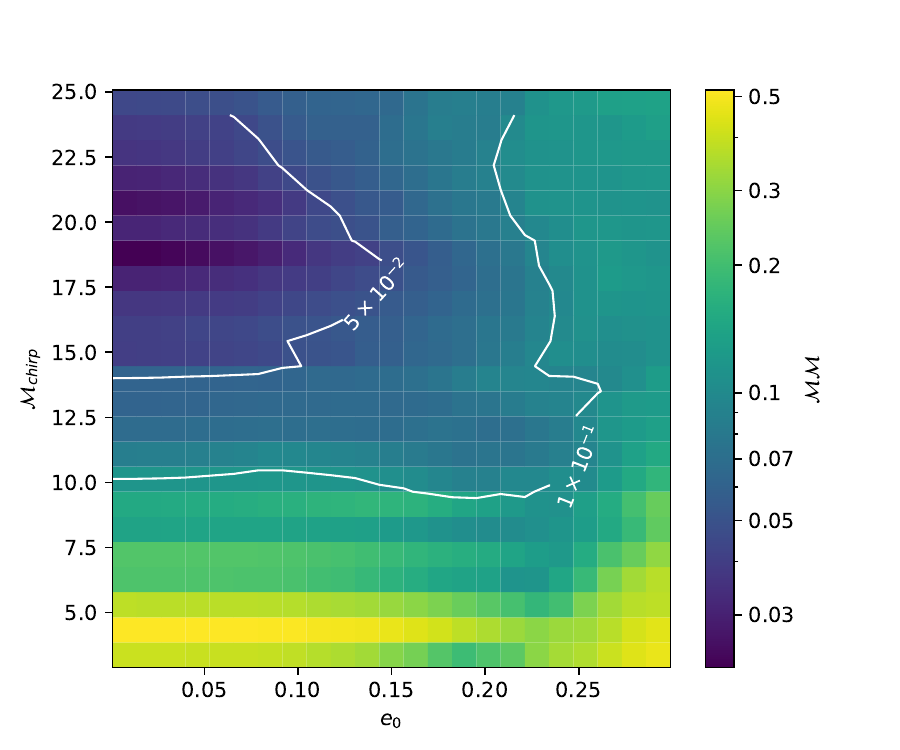}\hfill
  \includegraphics[width=0.49\textwidth]{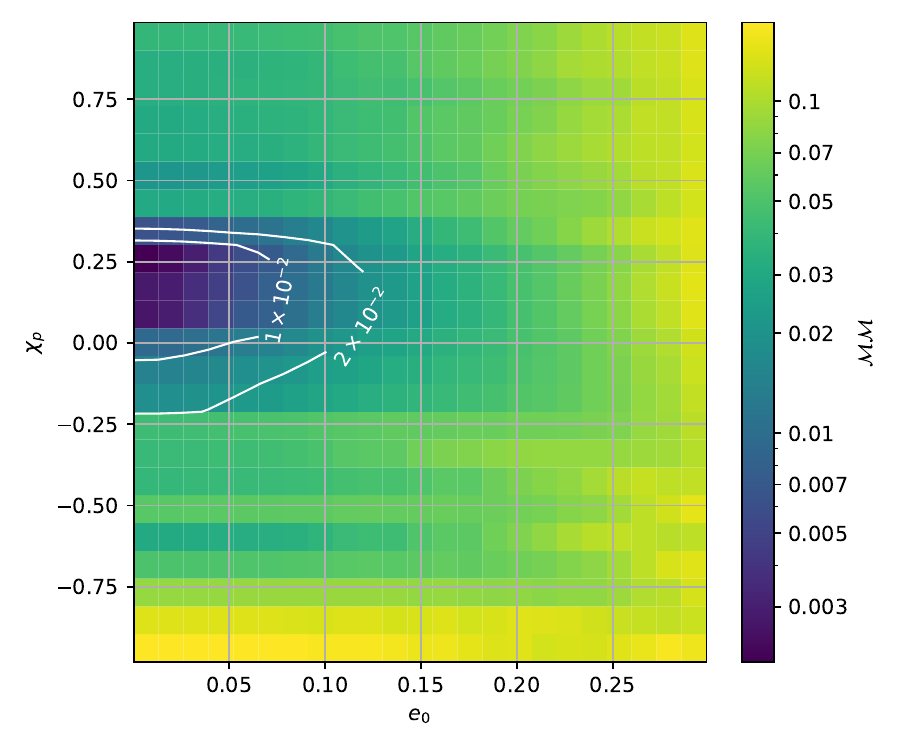}

  \caption{In the left figure, we set the initial spin configuration as $\chi_{1,x} = 0.6$, $\chi_{2,x} = 0.7$, $\chi_{1,z} = \chi_{2,z} = 0.2$, varied the chirp mass and initial eccentricity, and plotted the mismatch for various values of the same between \PFtwoEcc and \pyEFPE. As can be seen, for lower chirp masses which are of long duration, \PFtwoEcc accumulates more error given that the initial eccentricity accuracy order is $\mathcal{O}(e_0^2)$, and that it uses a Newtonian amplitude. However, the mismatch improves, and hence the usability is better for higher chirp masses. We also plot the $5\%$ and $10\%$ mismatch regions and notice that \PFtwoEcc performs best around $20\,M_{\odot}$. In the right figure, we fix the masses to be $20\,M_\odot$, $\chi_{1,z} = \chi_{2,z} = 0.2$, and varied the effective spin precession parameter $\chi_p$ and the initial eccentricity. We find that \PFtwoEcc performs better in the small eccentricity and small precession limit, whereas it performs poorly for high eccentricities and high precession. We also plot the $1\%$ and $2\%$ mismatch regions to illustrate the domain of validity of \PFtwoEcc.}
  \label{fig:TF2Ecc_and_pyEFPE_mismatch_side_by_side}
\end{figure*}
}

\section{Conclusions and Discussions}\label{sec:discussions}

In this work, we have developed, for the first time, closed-form analytical expressions that consistently capture the joint influence of orbital eccentricity and generic spin precession on the gravitational-wave (GW) phasing of compact binaries up to second post-Newtonian order. By employing the precession-averaging method of \cite{Morras:2025nlp}, we effectively removed the explicit time dependence in the spin–orbit and spin–spin interactions, thereby enabling tractable analytic solutions to the eccentricity evolution. Treating eccentricity as a small parameter, we obtained phasing formulae accurate up to the eighth power of the initial eccentricity $e_0$, and provided both time-domain (TaylorT2) and frequency-domain (TaylorF2/SUA) representations. Furthermore, we proposed a simple resummation procedure that extends the domain of validity of our TaylorT2 phase to eccentricities as high as $e_0 \sim 0.8$, thereby considerably enhancing the practical applicability of our results for data analysis. 

A key finding of our analysis is that even for binaries with nearly circular initial orbits, misaligned spins generically re-introduce small but non-negligible eccentricity during the inspiral. This highlights the intrinsic coupling between spin precession and eccentric orbital dynamics: while radiation reaction drives circularization, precession provides a persistent source of eccentricity. The resulting phase corrections accumulate over the GW cycles across the sensitivity bands of current and future detectors, underscoring the necessity of including both effects in template banks. Our comparison with numerical eccentricity evolution demonstrates that the analytically expanded and resummed phasing incurs less than a cycle of error for moderate eccentricities, well within the tolerances required for parameter estimation.

For binaries with almost aligned spins, eccentricity reduces the positive 1.5PN spin–orbit contributions and weakens the negative 2PN cycle subtraction, consistent with expectations that eccentricity depletes GW cycles in almost-aligned/aligned systems. In the almost anti-aligned case, reversing spin orientations flips the sign of the 1.5PN spin–orbit contributions, leading to a loss of cycles, while the 2PN spin–spin and higher-order couplings remain negative and decrease in magnitude with increasing eccentricity, mirroring the AA behavior. For the perpendicular configuration, the 1.5PN spin–orbit contributions vanish entirely, but the quadratic spin terms at 2PN persist, continuing to subtract cycles in a manner that predictably diminishes as eccentricity grows.

{\black A frequency domain Taylor waveform was also developed along with this study, which we call as \PFtwoEcc. In this waveform, developed in a private branch of \lal, we included our newly calculated eccentric and spin-precessing phasing terms up to 2 PN, as well as spin-aligned eccentric phasing terms up to 3 PN. Benchmark tests indicate a near perfect match with \FtwoEcc for zero spins, and high mismatch with the inclusion of spins. Similarly, a waveform comparison test with \pyEFPE revealed the domain of validity of \PFtwoEcc in various parameter spaces. We found that our waveform is valid for low eccentricity, high chirp mass, and low $\chi_p$, which is expected since our \Ftwo waveform only includes eccentricity corrections at the leading order, with no corrections beyond the Newtonian order (zero eccentricity) for the amplitude. Inclusion of higher order eccentricity terms in the \Ftwo waveform will improve the scenario drastically in the future, which however, is beyond the scope of the present work.}

From a data-analysis perspective, these results are especially timely. Recent claims of non-zero eccentricity in LIGO–Virgo–KAGRA detections reinforce the need for waveform models that faithfully combine eccentricity and spin-precession effects. Neglecting either effect risks introducing systematic biases in source parameter recovery, particularly in the estimation of spin orientations and eccentricities, which are key discriminants of binary formation channels. Our framework provides a computationally efficient route to build hybrid post-Newtonian and effective-one-body (EOB) models, and can also serve as a bridge to numerical relativity by supplying reliable analytic inputs at low to moderate eccentricities.

Looking forward, several avenues for further development remain. First, while the present treatment extends to 2PN order, including higher-order PN contributions—especially spin-precession corrections at 3PN and beyond—will be crucial for precision astrophysics. Second, a systematic calibration of our resummed phasing expressions against numerical relativity simulations of highly eccentric, precessing binaries will provide a robust test of their accuracy. Third, embedding these results within EOB and surrogate frameworks will allow efficient template generation suitable for real-time parameter estimation pipelines. Finally, the interplay of eccentricity and spin precession also opens a promising window for testing general relativity in dynamical, strong-field regimes.

In summary, our work represents an essential step toward building waveform models that simultaneously incorporate eccentricity and spin precession in a closed-form analytic framework. By improving both accuracy and computational efficiency, the results presented here are expected to significantly aid GW astronomy in the upcoming era of increasingly sensitive ground- and space-based detectors.

\acknowledgments

We are particularly grateful to Gonzalo Morras for clearing a significant number of queries during the course of this study, Divyajyoti for assistance with the review process, and Kaushik Paul for assistance with the mismatch analysis. We thank Patricia Schmidt for useful comments and suggestions on our manuscript. This document has LIGO preprint number LIGO-P2500628.

\appendix

\bibliographystyle{apsrev4-1}
%

\end{document}